\documentclass[
 reprint,
 amsmath,amssymb,
 aps,nofootinbib,preprintnumbers,longbibliography,superscriptaddress
]{revtex4-1}

\usepackage[utf8]{inputenc}
\usepackage{graphicx}
\usepackage{dcolumn}
\usepackage{bm}
\usepackage{hyperref}
\usepackage{lipsum}
\usepackage{xcolor}
\usepackage{calc}
\usepackage{accents}
\usepackage{comment}
\usepackage{float}
\usepackage{diagbox}
\usepackage{soul}
\usepackage{braket}
\usepackage[normalem]{ulem}

\font\BF=cmmib10
\def\k{{\hbox{\BF k}}}

\def\s{{\hbox{\BF s}}}

\def\r{{\hbox{\BF r}}}

\def\hinvMpc{h\,{\rm Mpc}^{-1}}
\def\Mpcinvh{{\rm Mpc}/h}
\newcommand{\code}[1]{\texttt{#1}}
\makeatletter
\newcommand\footnoteref[1]{\protected@xdef\@thefnmark{\ref{#1}}\@footnotemark}
\makeatother

\newcommand{\fedezc}{f_{\rm EDE}(z_c)}
\newcommand{\Planck}{{\it Planck}}
\newcommand{\lcdm}{$\Lambda$CDM}

\newcommand{\mnras}{MNRAS}

\begin{document}


\title{Updated constraints from the effective field theory analysis of BOSS power spectrum on Early Dark Energy}
\author{Th\'eo Simon}
\email{theo.simon@umontpellier.fr}
\affiliation{Laboratoire Univers \& Particules de Montpellier (LUPM), CNRS \& Universit\'e de Montpellier (UMR-5299), Place Eug\`ene Bataillon, F-34095 Montpellier Cedex 05, France}
\author{Pierre Zhang}
\email{pierrexyz@protonmail.com}
\affiliation{Department of Astronomy, School of Physical Sciences,
University of Science and Technology of China, Hefei, Anhui 230026, China \\
CAS Key Laboratory for Research in Galaxies and Cosmology,
University of Science and Technology of China, Hefei, Anhui 230026, China \\
School of Astronomy and Space Science, \\
University of Science and Technology of China, Hefei, Anhui 230026, China}
\author{Vivian Poulin}
\affiliation{Laboratoire Univers \& Particules de Montpellier (LUPM), CNRS \& Universit\'e de Montpellier (UMR-5299), Place Eug\`ene Bataillon, F-34095 Montpellier Cedex 05, France}
\author{Tristan L.~Smith}
\affiliation{Department of Physics and Astronomy, Swarthmore College, Swarthmore, PA 19081, USA}

\begin{abstract}
 Analyses of the full shape of BOSS DR12 power spectrum using the one-loop prediction from the effective field theory of large-scale structure (EFTBOSS) have led to new constraints on extensions to the $\Lambda$CDM model, such as early dark energy (EDE), which has been suggested as a resolution to the ``Hubble tension.'' 
 In this paper, we reassess the constraining power of the EFTBOSS on EDE in light of a correction to the normalization of BOSS window functions. Overall we find that constraints from EFTBOSS on EDE are weakened and represent a small change compared to constraints from \Planck~and the conventional BAO/$f\sigma_8$ measurements. The combination of \Planck{} data with EFTBOSS provides a bound on the maximal fractional contribution of EDE $f_{\rm EDE}<0.083$ at 95\% C.L. (compared to $<0.054$ with the incorrect normalization and $<0.088$ without full-shape data) and the Hubble tension is reduced to $2.1\sigma$. However, the more extreme model favored by an analysis with just data from the Atacama Cosmology Telescope is disfavored by the EFTBOSS data.
We also show that the updated Pantheon+ type Ia supernova (SN1a) analysis can slightly increase the constraints on EDE. Yet, the inclusion of the SN1a magnitude calibration by SH0ES strongly increases the preference for EDE to above $5\sigma$, yielding $f_{\rm EDE}\sim 0.12^{+0.03}_{-0.02}$ around the redshift $z_c=4365^{+3000}_{-1100}$.
Our results demonstrate that EFTBOSS data (alone or combined with \Planck~data) do not exclude the EDE resolution of the Hubble tension.
 \end{abstract}

\maketitle

\section{\label{sec:Intro}Introduction}

In recent years, several tensions between probes of the early and late Universe analyzed under the $\Lambda$ cold dark matter ($\Lambda$CDM) model have emerged.
The Hubble tension  refers to the inconsistency between local measurements of the current expansion rate of the Universe, \emph{i.e.},~the Hubble constant $H_0$, and the value inferred from early Universe data using the \lcdm\ model.
This tension is predominantly driven by the {\it Planck} Collaboration's observation of the cosmic microwave background (CMB), which predicts a value in \lcdm\ of $H_0 = 67.27 \pm 0.60$ km/s/Mpc \cite{Planck:2018vyg}, and the value measured by the SH0ES Collaboration using the Cepheid-calibrated cosmic distance ladder, whose latest measurement yields $H_0 = 73\pm1$ km/s/Mpc \cite{Riess:2021jrx,Riess:2022mme}. Taken at face value, these observations alone result in a $\sim 5\sigma$ tension.\footnote{A new calibration including cluster cepheids and Gaia EDR3 parallaxes further increase the tension to $5.3\sigma$ \cite{Riess:2022mme}.} 
Experimental efforts are underway to establish whether this discrepancy can be caused by yet unknown systematic effects (appearing in either the early or late Universe measurements \cite{Dainotti:2021pqg,Dainotti:2022bzg}, or both). 
 It appears that various attempts to alter the modeling of dust extinction are not successful in altering the Hubble constant \cite{Mortsell:2021nzg,Mortsell:2021tcx,Follin:2017ljs},  nor is there support for different populations of type Ia supernova (SNIa) at low$-z$ and high$-z$ causing significant impact \cite{Rigault:2014kaa,NearbySupernovaFactory:2018qkd,Jones:2018vbn,Brout:2020msh}.  
 In fact, the SH0ES team recently provided a comprehensive measurement of the $H_0$ parameter to 1.3\% precision, addressing these potential systematic errors, and concluded that there is ``{\em  no indication that the discrepancy arises
from measurement uncertainties or [over 70] analysis variations considered to date}'' \cite{Riess:2021jrx}. 
On the side of the CMB, it has been noted that \Planck{}~data carry a number of anomalies of low statistical significance that may play a role in this tension \cite{Addison:2015wyg,Planck:2016tof,Planck:2018vyg,DiValentino:2019qzk,Handley:2019tkm}.
 Nevertheless, the appearance of this discrepancy across an array of probes\footnote{For a very short summary of alternative methods, let us mention that, on the one hand, there exists a variety of different techniques for calibrating $\Lambda$CDM at high redshifts and subsequently inferring the value of $H_0$, which do not involve {\em Planck} data. For instance, one can use alternative CMB datasets such as WMAP, ACT, or SPT, or even remove observations of the CMB altogether and combine measurements of big bang nucleosynthesis (BBN)  with data from baryon acoustic oscillation (BAO) \cite{2019JCAP...10..029S,2018ApJ...853..119A}, resulting in $H_0$ values in good agreement with {\em Planck}.
On the other hand, alternative methods for measuring the local expansion rate have been proposed in the literature, in an attempt at removing any bias introduced from cepheid and/or SNIa observations. The Chicago-Carnegie Hubble program (CCHP), which calibrates SNIa using the tip of the red giant branch (TRGB), obtained a value of $H_0=69.8 \pm 0.6~\mathrm{(stat)} \pm 1.6~\mathrm{(sys)}$ km/s/Mpc \cite{Freedman:2019jwv,Freedman:2021ahq}, in between the {\em Planck} CMB prediction and the SH0ES calibration measurement, and a reanalysis of the CCHP data by Anand \emph{et al.} yields $H_0=71.5 \pm1.9$km/s/Mpc \cite{Anand:2021sum}. The SH0ES team, using the parallax measurement of $\omega-$Centauri from Gaia DR3 to calibrate the TRGB, obtained $H_0=72.1 \pm2.0$km/s/Mpc~\cite{Yuan:2019npk,Soltis:2020gpl}. Additional methods intended to calibrate SNIa at large distances include surface brightness fluctuations of galaxies \cite{Khetan:2020hmh}, Miras \cite{Huang:2019yhh}, or the Baryonic Tully Fisher relation \cite{Schombert:2020pxm}. There also exists a variety of observations that do not rely on observations of SNIa -- these include, \emph{e.g.}, time delay of strongly lensed quasars \cite{Wong:2019kwg,Birrer:2020tax}, maser distances \cite{Pesce:2020xfe}, or gravitational waves as ``standard sirens'' \cite{Abbott:2019yzh}.} (although not always with strong statistical significance) suggests that a single systematic effect may not be sufficient to resolve it. For recent reviews on the topic, we refer the reader to Refs.~\cite{DiValentino:2021izs,Abdalla:2022yfr}.

Additionally, within $\Lambda$CDM, the parameter  $S_8\equiv \sigma_8(\Omega_m/0.3)^{0.5}$, where $\sigma_8$ is the root mean square of matter fluctuations on an $8 h^{-1}$Mpc scale and $\Omega_m$ the (fractional) matter density today, inferred from CMB is about $2-3\sigma$ larger than that deduced from weak lensing surveys such as the CFHTLenS \cite{Heymans:2012gg}, KiDS-1000 \cite{Heymans:2020gsg}, DESY3 \cite{DES:2021wwk}, as well as from {\em Planck} Sunyaev-Zeldovich cluster abundances \cite{Planck:2018vyg, Planck:2015lwi} and SPT \cite{SPT:2018njh}. Additionally, the measurements of $S_8$ on large scales with galaxy clustering from BOSS full-shape data that have been reported also indicate a value that is on a low side, although not at an important significant level due to large error bars ($\sim 2\sigma$) \cite{Zhang:2021yna,Philcox:2021kcw}.~\footnote{Note that, however, these $S_8$ measurements might be affected by prior volume effects, as shown and quantified in~\cite{DAmico:2022osl}. Once those are accounted for, BOSS full-shape results and \Planck{} are brought to good agreement (see also~\cite{Amon:2022azi}).}
It is yet to be understood whether the $S_8$ tension is due to systematic effects \cite{Amon:2022ycy}, nonlinear modeling including the effect of baryons at very small scales \cite{Amon:2022azi}, or physics beyond $\Lambda$CDM.

Along with experimental developments to confirm the Hubble and $S_8$ tension, a lot of effort has been given to explain these discrepancies with some new physical mechanism, often in the form of extensions to the $\Lambda$CDM model that may be connected to the (still unknown) nature of dark matter or dark energy. 
It has been argued that the most promising category of solutions to resolve the $H_0$ tension involves physics in the pre-recombination era leading to a decrease of the sound horizon at recombination \cite{Bernal:2016gxb,Aylor:2018drw,Knox:2019rjx,Camarena:2021jlr,Efstathiou:2021ocp,Schoneberg:2021qvd}, such as models involving dark radiation and/or new neutrino properties \cite{Kreisch:2019yzn,Berbig:2020wve,Ghosh:2019tab,Forastieri:2019cuf,Escudero:2019gvw,Escudero:2021rfi,Blinov:2020hmc,Ghosh:2021axu,Archidiacono:2022ich,Aloni:2021eaq, Schoneberg:2022grr}, early dark energy (EDE) \cite{Karwal:2016vyq,Poulin:2018cxd,Smith:2019ihp,Niedermann:2019olb,Niedermann:2020dwg,Ye:2020btb,Poulin:2023lkg}, modified gravity \cite{Renk:2017rzu,Umilta:2015cta,Ballardini:2016cvy,Rossi:2019lgt,Braglia:2020iik,Zumalacarregui:2020cjh, Abadi:2020hbr,Ballardini:2020iws,Braglia:2020bym,DiValentino:2015bja,Bahamonde:2021gfp,Raveri:2019mxg,Yan:2019gbw,Frusciante:2019puu,SolaPeracaula:2019zsl,SolaPeracaula:2020vpg,Ballesteros:2020sik,Braglia:2020auw,Desmond:2019ygn,Lin:2018nxe}, or exotic recombination \cite{Chiang:2018xpn,Hart:2019dxi,Sekiguchi:2020teg,Jedamzik:2020krr,Cyr-Racine:2021alc} (for reviews, see Refs.~\cite{DiValentino:2021izs,Schoneberg:2021qvd}).

Interestingly, these models tend to leave signatures in the matter power spectrum on large scales that can be probed by large-scale structures surveys such as SDSS/BOSS~\cite{BOSS:2016wmc}. 
In fact, developments of the one-loop prediction of the galaxy power spectrum in redshift space from the effective field theory of large-scale structures (EFTofLSS)\footnote{See also the introduction footnote in, \emph{e.g.},~\cite{DAmico:2022osl} for relevant related works on the effective field theory of large-scale structures (EFTofLSS).}~\cite{Baumann:2010tm,Carrasco:2012cv,Senatore:2014via,Senatore:2014eva,Senatore:2014vja,Perko:2016puo}  have made possible the determination of the $\Lambda$CDM parameters from the full-shape analysis of SDSS/BOSS data~\cite{BOSS:2016wmc} at precision higher than that from conventional BAO and redshift space distortions (which measure the product $f\sigma_8$, where $f$ is the growth function) analyses, and even comparable to that of CMB experiments. 
This provides an important consistency test for the $\Lambda$CDM model, while allowing one to derive competitive constraints on models beyond $\Lambda$CDM (see, \emph{e.g.}, Refs.~\cite{DAmico:2019fhj,Ivanov:2019pdj,Colas:2019ret,DAmico:2020kxu,DAmico:2020tty,Simon:2022ftd,Chen:2021wdi,Zhang:2021yna,Philcox:2021kcw,Kumar:2022vee,Nunes:2022bhn,Lague:2021frh,Smith:2022iax,Simon:2022csv}). 
A thorough study of the consistency of EFTBOSS analyses within the $\Lambda$CDM model is presented in a companion paper \cite{Simon:2022lde}.

In this paper,  we reassess the constraints on EDE from the full shape of the most recent measurements of the power spectrum (or correlation function) of BOSS in light of a correction to the normalization of BOSS window functions (presented in App.~\ref{app:normalization}). 
EDE has been shown to reduce the Hubble tension to the $\sim1.5\sigma$ level, with an energy density representing at most a fraction $\fedezc\sim 12\%$ at the critical redshift $z_c\sim3500$ after which the fields start to dilute away  \cite{Karwal:2016vyq,Poulin:2018cxd,Smith:2019ihp,Schoneberg:2021qvd}. 
There exists a variety of other EDE models that can similarly reduce the tension to the $1.5-2.5\sigma$ level \cite{Lin:2019qug,Niedermann:2019olb,Berghaus:2019cls,Ye:2020btb,Karwal:2021vpk}. 
Recently, several groups have reported ``hints'' of EDE within ACT data at the $\sim3\sigma$ level, alone or in combination with WMAP (or, equivalently, \Planck{} temperature data restricted to $\ell < 650$) and \Planck{} polarization data \cite{Hill:2021yec,Poulin:2021bjr}, as well as with SPT-3G data \cite{LaPosta:2021pgm,Smith:2022hwi}. 

However, it has also been pointed out that EDE leaves an impact in the matter power spectrum that can be constrained thanks to the EFTofLSS applied to  BOSS data or through measurements of the parameter  $S_8$.
Typically, in the EDE cosmology that resolves the Hubble tension, the amplitude of fluctuations $\sigma_8$ is slightly larger due to increase in $\omega_{\rm cdm}$ and $n_s$, which are necessary to counteract some of the effects of the EDE on the CMB power spectra \cite{Poulin:2018cxd,Hill:2020osr,Vagnozzi:2021gjh}. As a result, the $S_8$ tension tends to increase by $\sim 0.5\sigma$ in the EDE cosmology, and large-scale structure (LSS) measurements may put pressure on the EDE model \cite{Hill:2020osr}. Additionally, it has been argued that the full-shape analysis of the galaxy power spectrum of BOSS disfavors the EDE model as an efficient resolution of the $H_0$ tension~\cite{Ivanov:2020ril,DAmico:2020ods}.
Indeed, in order to adjust the BAO data seen either in 3D or 2D at different comoving distances in a galaxy clustering survey (typically at $z\sim0.1-1$), it requires in the EDE cosmology an increase in $\omega_{\rm cdm}$\footnote{A similar increase is required to keep the CMB peaks' height fixed \cite{Poulin:2018cxd}, in particular, through the ISW effect \cite{Vagnozzi:2021gjh}.} \cite{Poulin:2018cxd,Jedamzik:2020krr}, which can affect the fit to the full-shape~\cite{Hill:2020osr,DAmico:2020ods,Ivanov:2020ril}. 
Thus, galaxy clustering data can provide a way to break the degeneracy introduced by EDE, in particular, due to the constraints it provides on $\omega_{\rm cdm}$ and $\sigma_8$. 

Although these effects are certainly relevant in constraining EDE, the original interpretation of the additional constraining power suggested in Refs.~\cite{DAmico:2020ods,Ivanov:2020ril} was disputed in Refs.~\cite{Smith:2020rxx,Murgia:2020ryi}. 
There, it was argued that the apparent constraining power from the BOSS full-shape analysis may be artificially amplified by (i) the impact of the prior volume artificially favoring $\Lambda$CDM in the Bayesian context (later verified with a profile likelihood approach\footnote{For further discussion about the mitigation of projection and prior volume effect, see Ref.~\cite{Gomez-Valent:2022hkb}.} \cite{Herold:2021ksg,Reeves:2022aoi}); (ii) a potential $\sim 20\%$ mismatch in the overall amplitude (typically parametrized by the primordial power spectrum amplitude $A_s$) between BOSS and \Planck, rather than additional constraints on $\omega_{\rm cdm}$. In parallel, it had already been pointed out in Ref.~\cite{Niedermann:2020qbw} that the effective field theory of LSS applied to BOSS data does not rule out the new EDE model.

In App.~\ref{app:normalization}, we explore the impact of the correction to the normalization of the BOSS data window function within $\Lambda$CDM and show that it leads to a $1\sigma$ shift upward in the value of $A_s$, now in better agreement with {\it Planck}.~\footnote{Note that, in our companion paper~\cite{Simon:2022lde}, we argue that the remaining difference on the amplitude might be explained by projection effects from the prior volume associated with the marginalization of the EFT parameters. }
Given that previous analyses, \emph{e.g.}, Refs.~\cite{DAmico:2020ods,Ivanov:2020ril}, have used the measurements inconsistently normalized between the power spectrum and the window function (as already acknowledged in Ref.~\cite{Philcox:2022sgj} for their previous analyses), the constraints from EDE are expected to change with these corrected BOSS measurements.
While Refs.~\cite{DAmico:2020ods,Ivanov:2020ril} concluded that the BOSS data, combined with {\it Planck} data, disfavored the EDE model as a potential candidate to solve the $H_0$ tension, we find here that the conclusions reached strongly depend on the normalization of the window functions used in the BOSS measurements.

Our paper is structured as follows: 
In Sec.~\ref{sec:modelanddata}, we review the EDE model and data considered in this work. In particular, we detail the possible choice of BOSS measurements and EFT likelihoods. 
In Sec.~\ref{sec:ede_eftoflss}, we assess the constraining power of corrected BOSS data alone on the EDE resolution to the Hubble tension and discuss differences between the constraints derived from the various BOSS data and EFTofLSS likelihoods.
In Sec.~\ref{sec:EFTCMB}, we derive constraints on EDE from the EFTBOSS data combined with either \Planck{} data (with and without SH0ES) or ACT data. We also show the impact of the new Pantheon+ SN1a catalog \cite{Brout:2022vxf} on the constraints on EDE.
We eventually present our conclusions in Sec.~\ref{sec:conclusions}.
App.~\ref{app:normalization} presents details on how to consistently normalize the window function with the power spectrum measurements.
App.~\ref{app:classpt_vs_pybird_EDE}, provides additional comparison between EFTofLSS likelihoods within the EDE model. 
Finally, App.~\ref{app:chi2} lists additional relevant information about $\chi^2$ statistics.

\section{Early Dark Energy Model and Data}
\label{sec:modelanddata}
\subsection{Brief review of the model}
The EDE model corresponds to an extension of the $\Lambda$CDM model, where the existence of an additional subdominant oscillating scalar field $\phi$ is considered. 
The EDE field dynamics is described by the Klein-Gordon equation of motion (at the homogenous level),
\begin{equation}
    \ddot{\phi} + 3 H \dot{\phi} + V_{n,\phi}(\phi) = 0\,,
\end{equation} 
where $V_n(\phi)$ is a modified axion-like potential defined as
\begin{equation}
    V_n(\phi) = m^2f^2\left[ 1- \cos(\phi/f)   \right]^n.
\end{equation}
$f$ and $m$ correspond to the decay constant and the effective mass of the scalar field, respectively, while the parameter $n$ controls the rate of dilution after the field becomes dynamical.
In the following, we will use the redefined field quantity $\Theta = \phi/f$ for convenience, such that $-\pi \le \Theta \le +\pi$.

At early times, when $H\gg m$, the scalar field $\phi$ is frozen at its initial value since the Hubble friction prevails, which implies that the EDE behaves like a form of dark energy and that its contribution to the total energy density increases relative to the other components. When the Hubble parameter drops below a critical value ($H \sim m$), the field starts evolving toward the minimum of the potential and becomes dynamical. The EDE contribution to the total budget of the Universe is maximum around a critical redshift $z_c$, after which the energy density starts to dilute with an approximate equation of state $w_{\phi} = P_{\phi}/\rho_{\phi}$~\cite{1983PhRvD..28.1243T,Poulin:2018dzj},
\begin{equation}
w_{\phi} = \left\{
    \begin{array}{ll}
        -1 \ &\text{if} \ z>z_c, \\
        \frac{n-1}{n+1} \ &\text{if} \ z<z_c.
    \end{array}
\right.
\end{equation}
In the following, we will fix $n=3$ as it was found that the data are relatively insensitive to this parameter provided $2\lesssim n \lesssim 5$ \cite{Smith:2019ihp}.
Instead of the theory parameters $f$ and $m$, we make use of $f_{\rm EDE}(z_c)$ and $z_c$ determined through a shooting method \cite{Smith:2019ihp}. We also include the initial field value $\Theta_i$ as a free parameter, whose main role once $f_{\rm EDE}(z_c)$ and $z_c$ are fixed is to set the dynamics of perturbations right around $z_c$, through the EDE sound speed $c_s^2$.

The EDE field will provide a small contribution to the expansion rate $H(z)$ around $z_c$ (we will focus on $\sim 10^3-10^4$ in the context of the Hubble tension), which causes a modification of the sound horizon at the recombination
\begin{equation}
    r_{\rm s}(z_{\rm rec}) = \int^{+\infty}_{z_{\rm rec}} \frac{c_s(z')}{H(z')}dz',
\end{equation}
where $c_s$ corresponds to the sound speed of the photon-baryon fluid acoustic waves. The sound horizon is observationally determined through the angular acoustic scale at recombination $\theta_s$, defined as
\begin{equation}
    \theta_s = \frac{r_s(z_{\rm rec})}{D_A(z_{\rm rec})} \ ,
\end{equation}
where $D_A(z_{\rm rec}) = \int_0^{z_{\rm rec}} dz'/H(z')\propto 1/H_0$  is the comoving angular diameter distance. Given that $\theta_s$ is determined from \Planck{} CMB power spectra with a very high accuracy, the change in the sound horizon must be compensated by a readjustment of the angular diameter distance in order to keep the angular acoustic scale constant. 
This readjustment is automatically done by increasing $H_0$ (and additional shift in $\omega_{\rm cdm}$ and $n_s$ to compensate effect of EDE on the growth of perturbations), which can, by design, bring the CMB measurements and the late-time estimate of the Hubble constant from the SH0ES Collaboration into agreement. 
In this paper, we address the question of whether the current full shape of galaxy clustering data analyzed using the EFTofLSS, can accommodate EDE. Indeed, on the one hand, the sound horizon seen at baryon-drag epoch $r_s(z_{\rm drag})$ is measured through another angular acoustic scale in galaxy surveys,
\begin{equation}
    \theta_g = \frac{r_s(z_{\rm drag})}{D_V(z_{\rm eff})} \ ,
\end{equation}
where $z_{\rm eff}$ is the effective redshift of the survey, and $D_V(z) = (D_A^2(z) \frac{c\cdot z}{H(z)})^{1/3}$ is a volume average of the comoving distances in the directions parallel and perpendicular to the line of sight, with $c$ the speed of light. The angle
$\theta_g$ typically summarizes the information from the BAO, and measuring it with high precision has the potential to break the degeneracy between $r_s(z_{\rm drag})$ and $H_0$ introduced by the EDE. 
In practice, BAOs from BOSS were shown to be well fit in combination with \Planck{} and SH0ES when allowing for EDE~\cite{Poulin:2018cxd}, at the cost of a larger  $\omega_{\rm cdm}$ \cite{Jedamzik:2020zmd}, which can simultaneously allow for the CMB peaks' height to be kept fixed \cite{Poulin:2018cxd} through the ISW effect \cite{Vagnozzi:2021gjh}.
However, the full-shape of the galaxy power spectrum also contains additional information. 
For example, the amplitude of the small-scale galaxy power spectrum at $k>k_{\rm eq}$,  where $k_{\rm eq}$ is the wavenumber entering the horizon at matter/radiation equality, contains information about $\omega_m$, $h$, and the spectral tilt $n_s$ \cite{DAmico:2019fhj,Colas:2019ret}. 
As the values of $\omega_{cdm}$ and $n_s$ are uplifted to compensate the growth of perturbations in the presence of EDE, the full shape of the galaxy power spectrum (with $\omega_b$ fixed by CMB or a BBN prior) is also modified in that respect. 
In the following, we quantify if these modifications from the EDE as a resolution of the $H_0$ tension are consistent with current cosmological data, including the full-shape galaxy power spectrum from BOSS modeled with the EFT.

\subsection{Data and method}\label{sec:data}

We analyze the EDE model in light of recent cosmological observations through a series of Markov chain Monte Carlo (MCMC) analyses using the Metropolis-Hastings algorithm from \texttt{MontePython-v3}\footnote{\url{https://github.com/brinckmann/montepython_public}.} code \cite{Audren:2012wb,Brinckmann:2018cvx} interfaced with our modified\footnote{\url{https://github.com/PoulinV/AxiCLASS}.} version of \texttt{CLASS}\footnote{\url{https://lesgourg.github.io/class_public/class.html}.} \cite{Blas:2011rf}. In this paper, we carry out various analyses from a combination of the following datasets:
\begin{itemize}
    \item {\bf PlanckTTTEEE:} The low-$l$ CMB TT, EE, and the high-$l$ TT, TE, EE data from \Planck{} 2018 \cite{Planck:2018vyg}.
    \item {\bf PlanckTT650TEEE:} Same dataset as \Planck{} TTTEEE, but in this case the TT power spectrum has a multipole range restricted to $l< 650$. 
    \item {\bf Lens:} The CMB gravitational lensing potential reconstructed from \Planck{} 2018 temperature and polarization data \cite{Planck:2018lbu}. When used without high-$l$ TT, TE, EE data, we use the CMB-marginalized version of the likelihood.\footnote{We thank Oliver Philcox for his help with correcting a bug in the standard Plik implementation.}
    \item {\bf ACT:}  The temperature and polarization angular power spectrum of the CMB from the Atacama Cosmology Telescope (ACT DR4) \cite{ACT:2020frw}. 
    \item {\bf BBN:} The BBN measurement of $\omega_b$ \cite{Schoneberg_2019} that uses the theoretical prediction of \cite{Consiglio_2018}, the experimental deuterium fraction of \cite{Cooke_2018}, and the experimental helium fraction of \cite{Aver_2015}.
    \item {\bf BAO:} The measurements of the BAO from the CMASS and LOWZ galaxy samples of BOSS DR12 at $z = 0.38$, 0.51, and 0.61 \cite{BOSS:2016wmc}, which we refer to as BOSS BAO DR12, and the BAO measurements from 6dFGS at $z = 0.106$ and SDSS DR7 at $z = 0.15$  \cite{Beutler:2011hx,Ross:2014qpa}, which we refer to as BOSS BAO low$-z$.
    \item {\bf BOSS $\bm{f\sigma_8}$ DR12:} We also sometimes include the redshift space distortion at $z = 0.38$, 0.51, and 0.61, which we refer to as $f\sigma_8$ \cite{BOSS:2016wmc}, taking into account the cross-correlation with BAO measurements.
    \item {\bf EFTBOSS:} The full-shape analysis of the BOSS power spectrum from the EFTofLSS, namely, $P_\textsc{fkp}^\textsc{lz/cm}$ \cite{Zhang:2021yna}, cross-correlated with reconstructed BAO, namely, $\alpha^\textsc{lz/cm}_\text{rec}$ \cite{Gil-Marin:2015nqa}.
    The measurements are defined in Table~\ref{tab:twopoint_summary}. 
    The SDSS-III BOSS DR12 galaxy sample data and covariances are described in Refs.~\cite{BOSS:2016wmc,Kitaura:2015uqa}. 
    The measurements, obtained in Ref.~\cite{Zhang:2021yna}, are from BOSS catalogs DR12 (v5) combined CMASS-LOWZ~\footnote{\url{https://data.sdss.org/sas/dr12/boss/lss/}.}~\cite{Reid:2015gra} and are divided in redshift bins LOWZ $0.2<z<0.43 \  (z_{\rm eff}=0.32)$ and CMASS $0.43<z<0.7  \ (z_{\rm eff}=0.57)$, with north and south Galactic skies for each, respectively, denoted NGC and SGC. 
    For the EDE analyses, we analyze the full shape of CMASS NGC, CMASS SGC, and LOWZ NGC, cross-correlated with post-reconstruction BAOs. 
    The analysis includes the monopole and quadrupole between $(k_{\rm min}, k_{\rm max}) = (0.01, 0.20/0.23) \hinvMpc$ in Fourier space and $(s_{\rm min}, s_{\rm max}) = (25/20, 200) \Mpcinvh$ in configuration space~\cite{Colas:2019ret,DAmico:2020kxu,Zhang:2021yna} for LOWZ / CMASS. 
    The theory prediction and likelihood are made available through \code{PyBird}.
    We also compare \code{PyBird} to \code{CLASS-PT}. More details on the differences between these likelihoods are given in Sec. II of Ref.~\cite{Simon:2022lde}. When computing constraints with \code{CLASS-PT}, we use the galaxy power spectrum monopole, quadrupole, and hexadecapole, for $0.01 \leqslant k \leqslant 0.2\ h{\rm Mpc}^{-1}$ as well as the real-space extension $Q_0$, up to $k_{\rm max} = 0.4\ h{\rm Mpc}^{-1}$, and the post-reconstructed BAO parameters. We use the standard \code{CLASS-PT} priors on the bias parameters.
    \item {\bf Pan18:} The Pantheon SNIa catalog, spanning redshifts $0.01 < z < 2.3$ \cite{Scolnic:2017caz}. We will also study in Sec.~\ref{sec:PanPlus} the impact of the newer Pantheon+ catalog, favoring a larger $\Omega_m$ \cite{Brout:2022vxf}, on our conclusions.
     \item {\bf SH0ES:} The SH0ES determination of $H_0 = 73.04\pm1.04$ km/s/Mpc from cepheid calibrated SNIa, modeled as a Gaussian likelihood.\footnote{For discussions about this modeling, see Refs.~\cite{Camarena:2021jlr,Efstathiou:2021ocp,Schoneberg:2021qvd}} 
\end{itemize}
 We will refer to the combination of \Planck TTTEEE+BAO+Pan18 as BaseTTTEEE, and to BaseTT650TEEE when replacing \Planck TTTEEE with \Planck TT650TEEE. In the absence of  CMB TTTEEE data, we refer to the dataset EFTBOSS+BBN+Lens+BAO+Pan18 as BaseEFTBOSS.
 For all runs performed, we use \textit{Planck} conventions for the treatment of neutrinos, that is, we include two massless and one massive species with $m_{\nu} = 0.06$ eV \cite{Planck:2018vyg}. In addition, we impose a large flat prior on the dimensionless baryon energy density $\omega_b$, the dimensionless cold dark matter energy density $\omega_{\rm cdm}$, the Hubble parameter today $H_0$, the logarithm of the variance of curvature perturbations centered around the pivot scale $k_p = 0.05$ Mpc$^{-1}$ (according to the \Planck{} convention), $\ln(10^{10}\mathcal{A}_s)$, the scalar spectral index $n_s$, and the reionization optical depth $\tau_{\rm reio}$. Regarding the three free parameters of the EDE model, we impose a logarithmic prior on $z_c$ and flat priors for $f_{\rm EDE}(z_c)$ and $\Theta_i$,
\begin{align*}
    &3 \le \log_{10}(z_c) \le 4, \\
    &0\le f_{\rm EDE}(z_c) \le 0.5, \\
    &0 \le \Theta_i \le \pi\,.
\end{align*}
We define our MCMC chains to be converged when the Gelman-Rubin criterion $R-1 < 0.05$, except for runs combining \Planck{}+EFTBOSS+ACT, for which we use a relaxed criterion of $R-1 < 0.1$ due to the complicated nature of the parameter space for the MCMC to explore.\footnote{Most parameters are converged at 0.01-0.05, the parameter with the worse convergence is $\theta_i$, which is often unconstrained or multimodal in the analyses.}
Finally, we extract the best-fit parameters from the procedure highlighted in the appendix of Ref.~\cite{Schoneberg:2021qvd}, and we produce our figures thanks to \code{GetDist} \cite{Lewis:2019xzd}.

\subsection{Details on the BOSS measurements and EFT likelihoods}
\label{sec:comparison_measurements}

\begin{table*}
\begin{tabular}{|l|l|l|l|l|l|}
\hline
\multicolumn{6}{|c|}{\bf Pre-reconstructed measurements}\\
\hline
        & Ref.       & Estimator       & Code                    & Redshift split     & Window             \\ \hline
$\mathcal{P}_\textsc{fkp}^\textsc{lz/cm}$              & \cite{Gil-Marin:2015sqa} & FKP  & \code{Rustico}\footnote{\url{https://github.com/hectorgil/Rustico}.}\cite{Gil-Marin:2015sqa}                 & LOWZ / CMASS       & Inconsistent norm. \\
$P_\textsc{fkp}^\textsc{lz/cm}$             & \cite{Zhang:2021yna} & FKP  & \code{PowSpec}\footnote{\url{https://github.com/cheng-zhao/powspec}.}~\cite{Zhao:2020bib} / \code{nbodykit}\footnote{\url{https://github.com/bccp/nbodykit}.}~\cite{Hand:2017pqn} & LOWZ / CMASS       & Consistent norm.   \\
$\xi^\textsc{lz/cm}$            & \cite{Zhang:2021yna} & Landy \& Slazay & \code{FCFC}\footnote{\url{https://github.com/cheng-zhao/FCFC}.}~\cite{Zhao:2020bib}                    & LOWZ / CMASS       & Window-free                 \\
$P_\textsc{fkp}^{z_1/z_3}$            & \cite{Beutler:2021eqq}~\footnote{\url{https://fbeutler.github.io/hub/deconv_paper.html}.}  & FKP  & --                       &  $z_1$ / $z_3$ & Consistent norm.   \\
$P_\textsc{quad}^{z_1/z_3}$            & \cite{Philcox:2021kcw} & Quadratic       & \code{Spectra without Windows}\footnote{\url{https://github.com/oliverphilcox/Spectra-Without-Windows}.}~\cite{Philcox:2020vbm} &  $z_1$ / $z_3$ & Window-free                 \\ \hline  
\multicolumn{6}{c}{}\\[-0.5em]
\hline
\multicolumn{6}{|c|}{\bf Post-reconstructed measurements}\\
\hline
& Ref. & -- & -- & Redshift split & Method \\ \hline
$\alpha^\textsc{lz/cm}_\text{rec}$           & \cite{Gil-Marin:2015nqa} & --               & --                       & LOWZ / CMASS       & \cite{DAmico:2020kxu}                 \\
$\alpha^{z_1/z_3}_\text{rec}$           & \cite{BOSS:2016hvq} & --               & --                       &  $z_1$ / $z_3$ &  \cite{DAmico:2020kxu}               \\
$\beta^{z_1/z_3}_\text{rec}$            & \cite{BOSS:2016hvq} & --               & --                       &  $z_1$ / $z_3$ & \cite{Philcox:2020vvt}                  \\ \hline
\end{tabular}
\caption{Comparison of pre and post-reconstructed BOSS two-point function measurements: reference, estimator, code of the measurements, redshift split [LOWZ, $0.2<z<0.43 \  (z_{\rm eff}=0.32)$; CMASS, $0.43<z<0.7  \ (z_{\rm eff}=0.57)$; $z_1$, $0.2<z<0.5 \  (z_{\rm eff}=0.38)$;  $z_3$: $0.5<z<0.7  \ (z_{\rm eff}=0.61)$], and window function treatment. 
For the post-reconstructed measurements, while we instead provide under ``Method'' the references presenting the algorithm used to extract the reconstructed BAO parameters and how the cross-correlation with the pre-reconstructed measurements is performed, ``Ref.'' now refers to the public post-reconstructed measurements used.
The SDSS-III BOSS DR12 galaxy sample data are described in Refs.~\cite{BOSS:2016wmc,Kitaura:2015uqa}. 
The pre-reconstructed measurements are from BOSS catalogs DR12 (v5) combined CMASS-LOWZ~\footnote{\url{https://data.sdss.org/sas/dr12/boss/lss/}.}~\cite{Reid:2015gra}. More details can be found in Sec. IV of Ref.~\cite{Simon:2022lde}. 
}
\label{tab:twopoint_summary}
\end{table*}

In this paper, we perform a thorough comparison of the constraints derived from the EFTBOSS data, in order to assess the consistency of the various analyses presented in the literature.  
Indeed, there are various BOSS two-point function measurements available to perform full-shape analyses, as well as a different EFT code.
As described in more detail in Ref.~\cite{Simon:2022lde}, the BOSS DR12 data can be divided into two different sets of redshift splitting (LOWZ/CMASS vs $z_1$/$z_3$).  
Furthermore, depending on the estimator, the data are sometimes analyzed by convolving the theory model with a window function, or not. For a window-free analysis, one way is to use the configuration-space correlation function $\xi$, another is to use a quadratic estimator, which we denote with the subscript ``QUAD.'' Finally, there are different ways to analyze the post-reconstructed parameters, which are then combined with the EFTBOSS data, denoted by $\alpha_{\rm rec}$ and $\beta_{\rm rec}$. These different datasets include slightly different amounts of information (due to different scale cuts) but they all represent reasonable choices on how to analyze the BOSS DR12 observations.

The characteristics of each measurement are listed in Tab.~\ref{tab:twopoint_summary} and more details can be found in Sec.~IV of Ref.~\cite{Simon:2022lde}. 
The EFT implementation and BOSS data we will focus on in this study are packaged in the \code{PyBird} likelihood, based on the EFT prediction and likelihood from~\code{PyBird}~\footnote{\url{https://github.com/pierrexyz/pybird}.}~\cite{DAmico:2020kxu}, and the \code{CLASS-PT} likelihood,  based on the EFT prediction from~\code{CLASS-PT}~\footnote{\url{https://github.com/michalychforever/CLASS-PT}.}~\cite{Chudaykin:2020aoj} and likelihood from Ref.~\cite{Philcox:2021kcw}.~\footnote{\url{https://github.com/oliverphilcox/full_shape_likelihoods}.}
Details about the \code{PyBird} and \code{CLASS-PT} likelihoods are presented in Sec.~II of Ref.~\cite{Simon:2022lde}.
Here, let us simply mention that \code{CLASS-PT} implements the IR-resummation scheme proposed in Ref.~\cite{Blas:2016sfa} and generalized to redshift space in Ref.~\cite{Ivanov:2018gjr}. This is different than that implemented in \code{PyBird}, proposed in Ref.~\cite{Senatore:2014via}, generalized to redshift space in Ref.~\cite{Lewandowski:2015ziq}, and made numerically efficient in Ref.~\cite{DAmico:2020kxu}. 
The \code{CLASS-PT} scheme has been shown to be an approximation of the one used in \code{PyBird} in Ref.~\cite{Lewandowski:2018ywf}, where one considers only the resummation of the bulk displacements around the BAO peak, $r_{\rm BAO} \sim 110 \Mpcinvh$.
For this scheme to be made practical, one further relies on a wiggle-no-wiggle split procedure to isolate the BAO part. 
Although this scheme has been shown to work fairly well within $\Lambda$CDM for cosmologies not too far from the one of \Planck, we cautiously observe that in far-away cosmologies as the ones probed in EDE, the BAO peak location happens to be dramatically modified, and it thus remains to be checked that the approximations still hold in these cases. 
For our prior choice (on $f_{\rm EDE}$), we have checked that at least the wiggle-no-wiggle split procedure as implemented in \code{CLASS-PT} is as numerically stable as for a fiducial case where the BAO peak is $\sim 110 \Mpcinvh$. 

In addition, in Ref.~\cite{Simon:2022lde}, we have checked the validity of the two pipelines by implementing in the \code{PyBird} likelihood the exact same prior as those used in the \code{CLASS-PT} likelihood, and we found agreement on the 1D posteriors of the cosmological parameters at $\lesssim 0.2\sigma$ in $\Lambda$CDM, where these residual differences can be attributed to the different implementations of the IR-resummation mentioned above.

\section{Updated EFTBOSS constraints on EDE} 
\label{sec:ede_eftoflss}

\subsection{Preliminary study}

\begin{figure*}
\centering
\includegraphics[trim=1cm 0cm 3cm 1cm, clip=true,width=2\columnwidth]{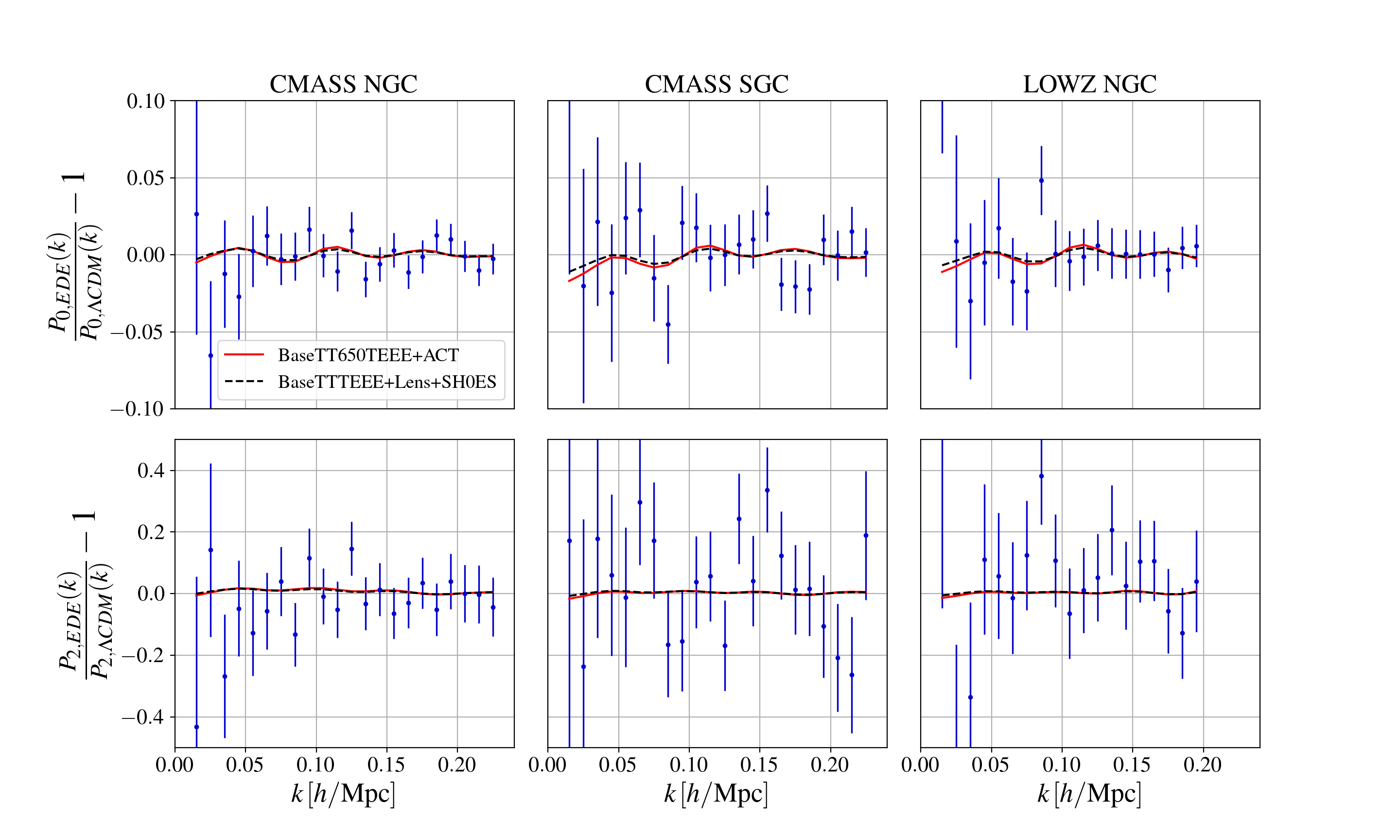}
\caption{Residuals of the monopole and quadrupole of the galaxy power spectrum in two EDE models (see. Tab.~\ref{tab:EDE_preliminary}) with respect to the $\Lambda$CDM model (obtained from the baseTTTEEE+Lens+EFTBOSS analysis \cite{Simon:2022ftd}) for the three sky cuts of the EFTBOSS data. }
\label{fig:galaxy_power_spectrum_prliminary}
\end{figure*}

In the recent literature, there has been a number of analyses showing hints of EDE and allowing for a resolution of the Hubble tension \cite{Poulin:2018cxd,Niedermann:2019olb,Hill:2021yec,Poulin:2021bjr,LaPosta:2021pgm,Smith:2022hwi}. In this preliminary study, we will take the results of two representative analyses. First, the baseline analysis of BaseTTTEEE+Lens+SH0ES data (second column of Tab.~\ref{tab:cosmoparam_planck}) has a best-fit of $f_{\rm EDE}(z_c) = 0.122$, $H_0 = 71.89$ km.s$^{-1}$.Mpc$^{-1}$.  Second, the analysis of BaseTT650TEEE+ACT (first column of Tab.~\ref{tab:cosmoparam_act}) favors an EDE model with significantly larger values of $f_{\rm EDE}(z_c)$ and $H_0$ compared to the BaseTTTEEE+Lens+SH0ES, namely, $f_{\rm EDE}(z_c) = 0.159$, $H_0 = 73.30$ km.s$^{-1}$.Mpc$^{-1}$ (see also \cite{Hill:2021yec,LaPosta:2021pgm,Poulin:2021bjr,Smith:2022hwi}). 
In this section, we will gauge how these two specific models fair against BOSS data following Refs.~\cite{DAmico:2020ods,Ivanov:2020ril}.

Using the best-fit parameters listed in  Tab.~\ref{tab:cosmoparam_planck} (second column) and Tab.~\ref{tab:cosmoparam_act} (first column), we perform a preliminary study where we determine the $\chi^2$ of the EFTBOSS data (using our fiducial $P_\textsc{fkp}^\textsc{lz/cm}+\alpha^\textsc{lz/cm}_\text{rec}$ data) after optimizing only the EFT parameters (since the cosmological parameters are fixed here). 
Using the \code{PyBird} code, we show in Tab.~\ref{tab:EDE_preliminary} the $\chi^2$ associated with the EFTBOSS data, and we plot in Fig.~\ref{fig:galaxy_power_spectrum_prliminary} the residuals with respect to $\Lambda$CDM from the BaseTTTEEE+Lens+EFTBOSS analysis\footnote{When combined with EFTBOSS, we do not include the BOSS BAO+$f\sigma_8$ data. } \cite{Simon:2022ftd}. We also show the BOSS data residuals for comparison with respect to the same model.   First, one can see that the changes in the residuals between those various fits are almost imperceptible by eye with respect to BOSS error bars. 
We find that the $\chi^2$ of the BOSS data is degraded by $+1.1$ for BaseTTTEEE+Lens+SH0ES (to be compared with $\sim+2.5$ in Ref.~\cite{Ivanov:2020ril}) and $+2.4$ for BaseTT650TEEE+ACT, compared to the best-fit $\chi^2$ of EFTBOSS data in the $\Lambda$CDM model. 
Despite this small $\chi^2$ degradation, we note that the $p$-value of BOSS data in the EDE models that resolve the Hubble tension is still very good.
Nevertheless, we anticipate that the EFTBOSS data could have a non-negligible constraining power in combination with BaseTT650TEEE+ACT, while its impact should be small in the context of the  BaseTTTEEE+Lens+SH0ES analysis. 

\begin{table*}[t]
\begin{tabular}{|l|c|c|c|}
 \hline
  & BaseTTTEEE+Lens &  BaseTT650TEEE & BaseTTTEEE+Lens\\
  & +SH0ES (EDE) & +ACT (EDE)& +EFTBOSS ($\Lambda$CDM)\\
\hline
$\chi^2_{\text{CMASS NGC}}$ & 39.3 & 39.1 & 40.3 \\
$\chi^2_{\text{CMASS SGC}}$ & 45.2 & 46.0 & 44.0  \\
$\chi^2_{\text{LOWZ NGC}}$ & 34.4 & 35.1 & 33.5 \\
\hline
$\chi^2_{\text{EFTBOSS}}$ & 118.9 & 120.2 & 117.8 \\
 $\Delta\chi^2_{\rm min} (\text{EDE}-\Lambda\text{CDM})$ &  +1.1 & + 2.4 & -- \\
\hline
$p-$value ($\%$) & $16.7$ & $14.7$ & $18.5$ \\
\hline
$N_{\rm data}$ &\multicolumn{3}{|c|}{132}\\
\hline
\end{tabular}
\caption{$\chi^2$ of each sky cut of the EFTBOSS dataset for the EDE best-fit models extracted from a fit to BaseTTTEEE+Lens+SH0ES and BaseTT650TEEE+ACT and the $\Lambda$CDM model from a fit to BaseTTTEEE+Lens+EFTBOSS. 
We also indicated the $\Delta\chi^2$ with respect to the $\Lambda$CDM best-fit model. 
The associated $p$-value is calculated assuming that the data points are uncorrelated and taking $3 \cdot 9$ EFT parameters in each fit (given that the cosmology is fixed). }
\label{tab:EDE_preliminary}
\end{table*}

\subsection{Constraints from various BOSS data}
\label{sec:EDE_arena_main}

\begin{figure}[t!]
\centering
\includegraphics[width=1.\columnwidth]{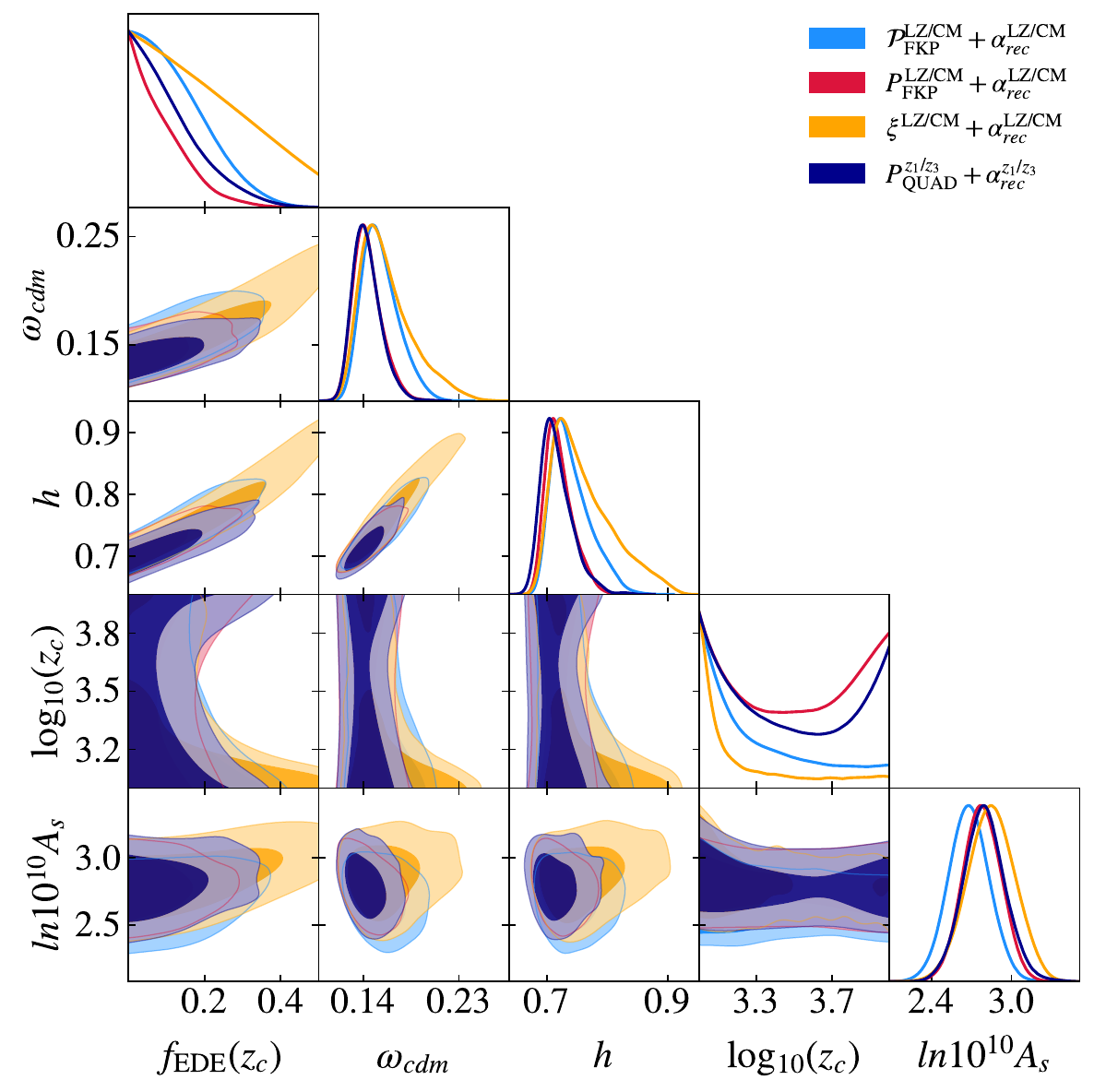}
\caption{Comparison of 2D posteriors of a subset of parameters in the EDE model reconstructed from BOSS full-shape analyses using \code{PyBird} baseline likelihood, with a BBN prior on $\omega_b$, of various pre-reconstructed two-point function measurements and handling of the window functions ($\mathcal{P}_\textsc{fkp}^\textsc{lz/cm}, P_\textsc{fkp}^\textsc{lz/cm}, \xi^\textsc{lz/cm}, P_\textsc{quad}^{z_1/z_3}$) combined with various post-reconstructed BAO parameters ($\alpha^\textsc{lz/cm}_\text{rec}$, $\alpha^{z_1/z_3}_\text{rec}$).
We recall that $\mathcal{P}_\textsc{fkp}^\textsc{lz/cm}+\alpha^\textsc{lz/cm}_\text{rec}$ corresponds to the BOSS FKP measurements analyzed with the EFT predictions convolved with inconsistently normalized window functions.
The main EDE analyses of this work are based on EFTBOSS, which corresponds to $P_\textsc{fkp}^\textsc{lz/cm} + \alpha^\textsc{lz/cm}_\text{rec}$. 
We choose to show only the cosmological parameters that are not completely prior dominated.}
\label{fig:boss_ede_summary_triangle}
\end{figure}

\begin{figure*}
\centering
    {\large \emph{Constraints from BOSS+BBN on EDE}}\\[0.2cm]
\includegraphics[width=2.\columnwidth]{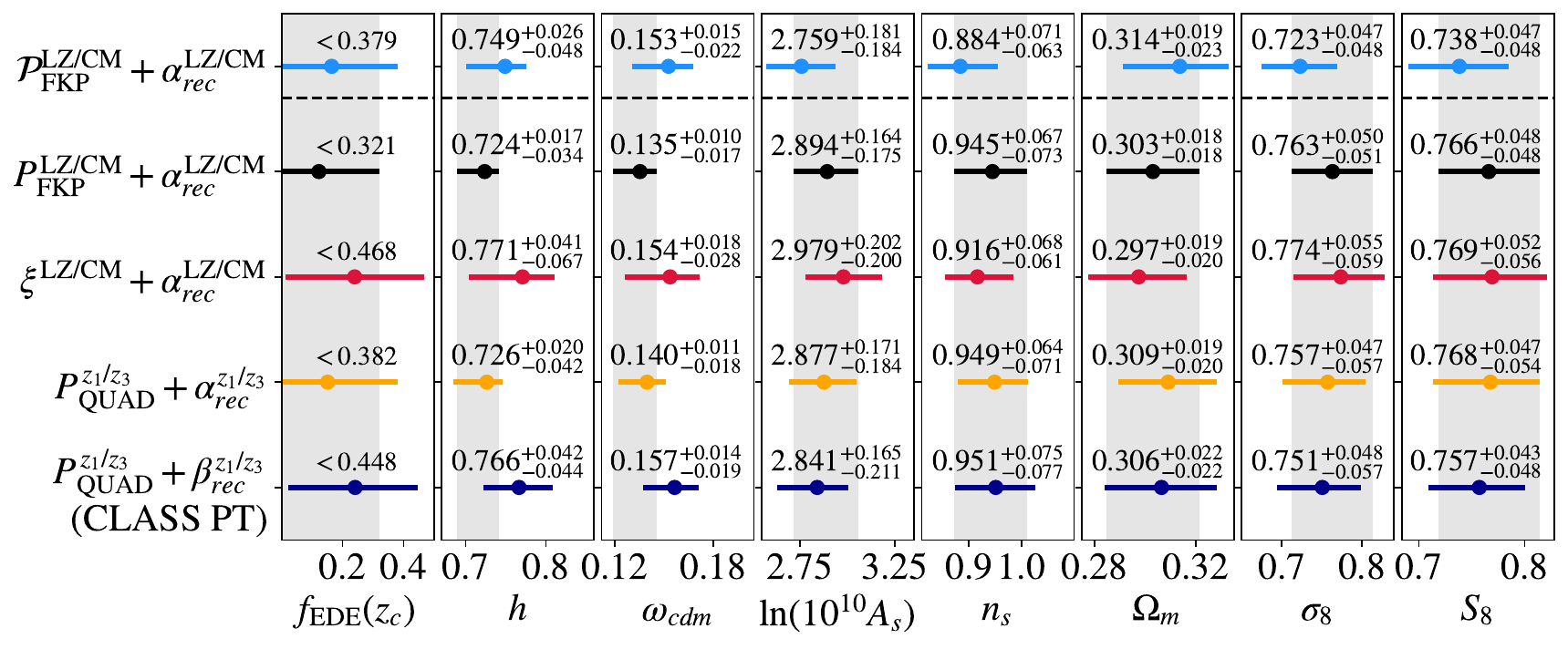}
\caption{Comparison of 1D credible intervals in the EDE model reconstructed from BOSS full-shape analyses using \code{PyBird} baseline likelihood, with a BBN prior on $\omega_b$, of various pre-reconstructed two-point function measurements and handling of the window functions ($\mathcal{P}_\textsc{fkp}^\textsc{lz/cm}, P_\textsc{fkp}^\textsc{lz/cm}, \xi^\textsc{lz/cm}, P_\textsc{quad}^{z_1/z_3}$) combined with various post-reconstructed BAO parameters ($\alpha^\textsc{lz/cm}_\text{rec}, \alpha^{z_1/z_3}_\text{rec}$, and $\beta^{z_1/z_3}_\text{rec}$).
We recall that $\mathcal{P}_\textsc{fkp}^\textsc{lz/cm}+\alpha^\textsc{lz/cm}_\text{rec}$ corresponds to the BOSS FKP measurements analyzed with the EFT predictions convolved with inconsistently normalized window functions.
The gray region corresponds to the EFTBOSS data that we use in our main analysis, namely, $P_\textsc{fkp}^\textsc{lz/cm} + \alpha^\textsc{lz/cm}_\text{rec}$.
In the last line, we also show the results of $P_\textsc{quad}^{z_1/z_3}+\beta^{z_1/z_3}_\text{rec}$ analyzed using the \code{CLASS-PT} baseline likelihood. 
Relevant information regarding the measurements and their notations are summarized in Tab.~\ref{tab:twopoint_summary}.
We choose to show only the cosmological parameters that are not prior dominated. 
For $f_{\rm EDE}$, we quote instead the $2\sigma$ bound.}
\label{fig:boss_ede_summary}
\end{figure*}

As is done in Ref.~\cite{Simon:2022lde} for $\Lambda$CDM, we compare the constraints on EDE from the various BOSS two-point function measurements, described in Tab.~\ref{tab:twopoint_summary}, in combination with the BBN prior on $\omega_b$.

The comparison of the 2D posteriors is shown in Fig.~\ref{fig:boss_ede_summary_triangle}, while the 1D posteriors of $\{f_{\rm EDE}(z_c), h,\omega_{\rm cdm},\ln(10^{10}A_s),n_s,\Omega_m\,\sigma_8,S_8\}$ are shown in Fig.~\ref{fig:boss_ede_summary}. 
In these figures, we also display the results from the BOSS data analyzed with the EFT predictions convolved with inconsistently normalized window functions, namely, $\mathcal{P}_\textsc{fkp}^\textsc{lz/cm}+\alpha^\textsc{lz/cm}_\text{rec}$, which disfavor the EDE model when they are combined with {\em Planck} data \cite{DAmico:2020ods,Ivanov:2020ril} (see the discussion in App.~\ref{app:normalization} for the impact of inconsistent normalization within the $\Lambda$CDM model). 
Interestingly, using the \code{PyBird} likelihood, the $\Lambda$CDM parameters are broadly consistent between $P_\textsc{fkp}^\textsc{lz/cm}+\alpha^\textsc{lz/cm}_\text{rec}$ and $P_\textsc{quad}^{z_1/z_3}+\alpha^{z_1/z_3}_\text{rec}$, as we have a shift of $\lesssim 0.3\sigma$ on $\Lambda$CDM parameters between these two measurements.
However, we find that $P_\textsc{fkp}^\textsc{lz/cm}+\alpha^\textsc{lz/cm}_\text{rec}$ leads to stronger constraints on EDE, namely,\footnote{Per convention, we cite one-sided bound at 95\% C.L. and two-sided ones at 68\% C.L.} $f_{\rm EDE}(z_c) < 0.321$, while $P_\textsc{quad}^{z_1/z_3}+\alpha^{z_1/z_3}_\text{rec}$ yields $f_{\rm EDE}(z_c) < 0.382$.

Concerning $\xi^\textsc{lz/cm}+\alpha^\textsc{lz/cm}_\text{rec}$, we find different constraints, even for the $\Lambda$CDM parameters: comparing $\xi^\textsc{lz/cm}+\alpha^\textsc{lz/cm}_\text{rec}$ to $P_\textsc{fkp}^\textsc{lz/cm}+\alpha^\textsc{lz/cm}_\text{rec}$, we find shifts of $\lesssim 1.2\sigma$, whereas comparing $\xi^\textsc{lz/cm}+\alpha^\textsc{lz/cm}_\text{rec}$ to $P_\textsc{quad}^{z_1/z_3}+\alpha^{z_1/z_3}_\text{rec}$, we find shifts of $\lesssim 1.0\sigma$. Let us note that the constraints on $\Lambda$CDM parameters reconstructed from $\xi^\textsc{lz/cm}+\alpha^\textsc{lz/cm}_\text{rec}$ are weaker than those of $P_\textsc{fkp}^\textsc{lz/cm}+\alpha^\textsc{lz/cm}_\text{rec}$ and $P_\textsc{quad}^{z_1/z_3}+\alpha^{z_1/z_3}_\text{rec}$, which is consistent with what was found within the $\Lambda$CDM model in our companion paper~\cite{Simon:2022lde} (see also Ref.~\cite{Zhang:2021yna} and explanations therein).
Regarding the EDE parameters, we obtain weaker constraints on $f_{\rm EDE}$, namely $f_{\rm EDE}(z_c) < 0.468$.
It is worth noting that, for the same likelihood, the constraints on $f_{\rm EDE}(z_c)$ can be up to 
$\sim35\%$ different depending on the data (especially between $P_\textsc{fkp}^\textsc{lz/cm} + \alpha^\textsc{lz/cm}_\text{rec}$ and $\xi^\textsc{lz/cm} + \alpha^\textsc{lz/cm}_\text{rec}$).
However, regardless of the data we consider, the BOSS full-shape (analyzed on their own with a BBN prior) within EDE leads to reconstructed values of $H_0$ that are compatible with what is obtained by the SH0ES Collaboration.

This conclusion also holds for the \code{CLASS-PT} baseline (last line of Fig.~\ref{fig:boss_ede_summary}), which is less constraining than the \code{PyBird} likelihood for the EDE model.  Indeed, we obtain $f_{\rm EDE}(z_c) < 0.448$, which is $\sim 15\%$ weaker than the constraint obtained with the \code{PyBird} likelihood, even for similar data ($P_\textsc{quad}^{z_1/z_3}$).
Furthermore, we note that the $f_{\rm EDE}(z_c)$ constraint reconstructed from $P_\textsc{fkp}^\textsc{lz/cm} + \alpha^\textsc{lz/cm}_\text{rec}$,  analyzed with the \code{PyBird} likelihood, is $\sim35\%$ weaker than the constraint obtained from $P_\textsc{quad}^{z_1/z_3}+\beta^{z_1/z_3}_\text{rec}$, analyzed with the \code{CLASS-PT} likelihood. 
We conclude that the standard \code{PyBird} analysis setup (which consists of our baseline setup) shows a higher constraining power than the standard \code{CLASS-PT} analysis.
Let us note that, for the $H_0$ parameter, we obtain a value 1.4$\sigma$ higher than the \Planck{} value ($h = 0.6851_{-0.014}^{+0.0076}$ at $68\%$ CL) with the \code{PyBird} analysis setup, and a value 1.8$\sigma$ higher with the \code{CLASS-PT} analysis setup, which indicates a reasonably good consistency between \Planck{} and BOSS regarding $H_0$.
For a more detailed discussion, including other data combinations, of the differences between \code{PyBird} and \code{CLASS-PT} for the EDE model, we refer to App.~\ref{app:classpt_vs_pybird_EDE}.
We, however, warn that the cosmological constraints from EFTBOSS at the level of the 1D posteriors might be affected by prior effects, as discussed in our companion paper~\cite{Simon:2022lde} in the context of $\Lambda$CDM.

\subsection{Primary CMB-free constraints on EDE}

\begin{figure*}
    \centering
        \centering
    \includegraphics[width=2\columnwidth]{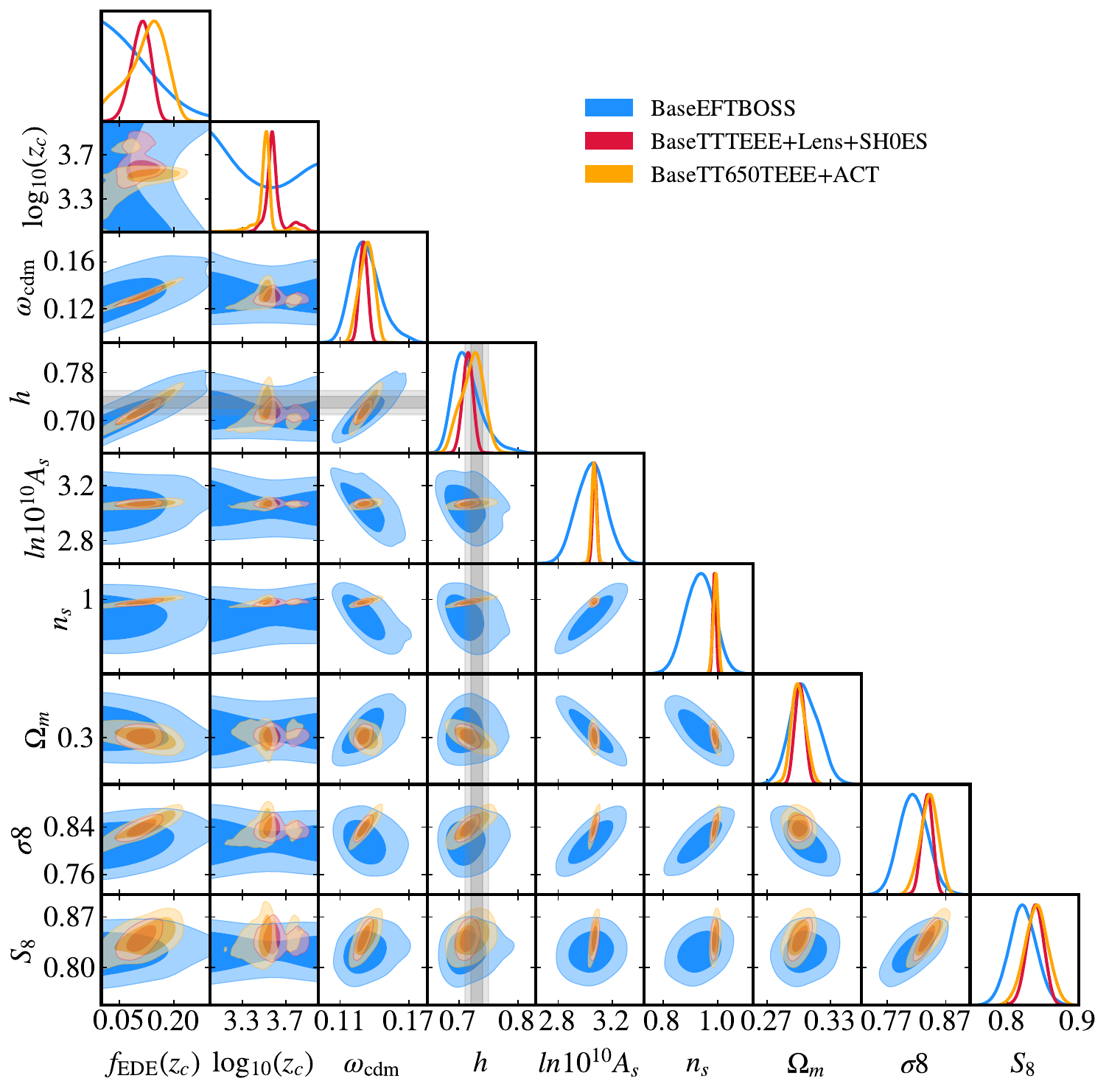}
    \caption{2D posterior distributions reconstructed from the BaseEFTBOSS dataset compared with the posterior reconstructed from BaseTTTEEE+Lens+SH0ES and BaseTT650TEEE+ACT.  We recall that BaseEFTBOSS refers to EFTBOSS+BBN+Lens+BAO+Pan18, BaseTTTTEEE refers to \Planck TTTEE+BAO+Pan18, and   BaseTT650TEEE to \Planck TT650TEE+BAO+Pan18.
   }
    \label{fig:EDE_EFT}
\end{figure*}

\begin{figure*}
    \centering
        \centering

    \includegraphics[width=1.\columnwidth]{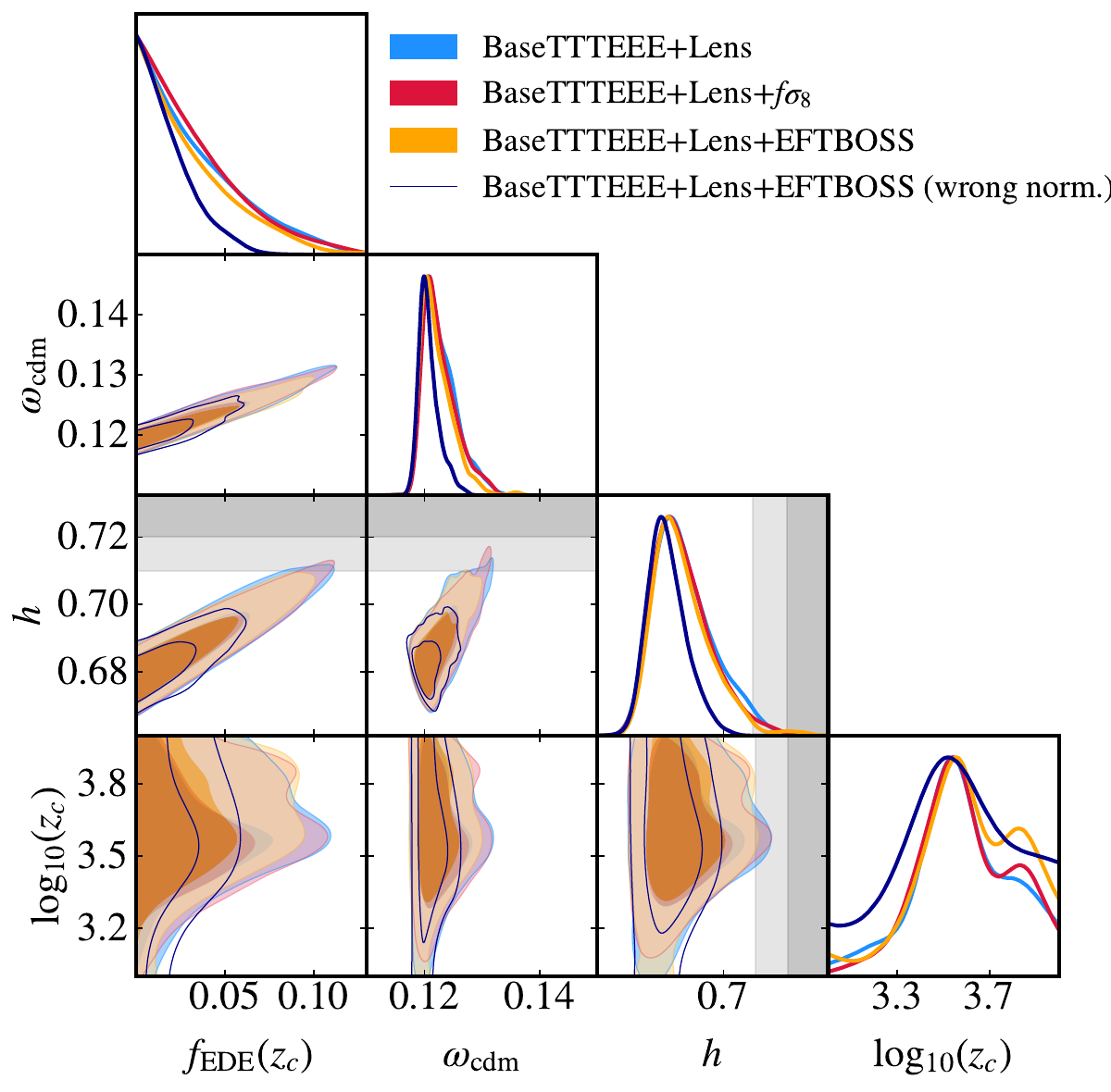}
    \includegraphics[width=1.\columnwidth]{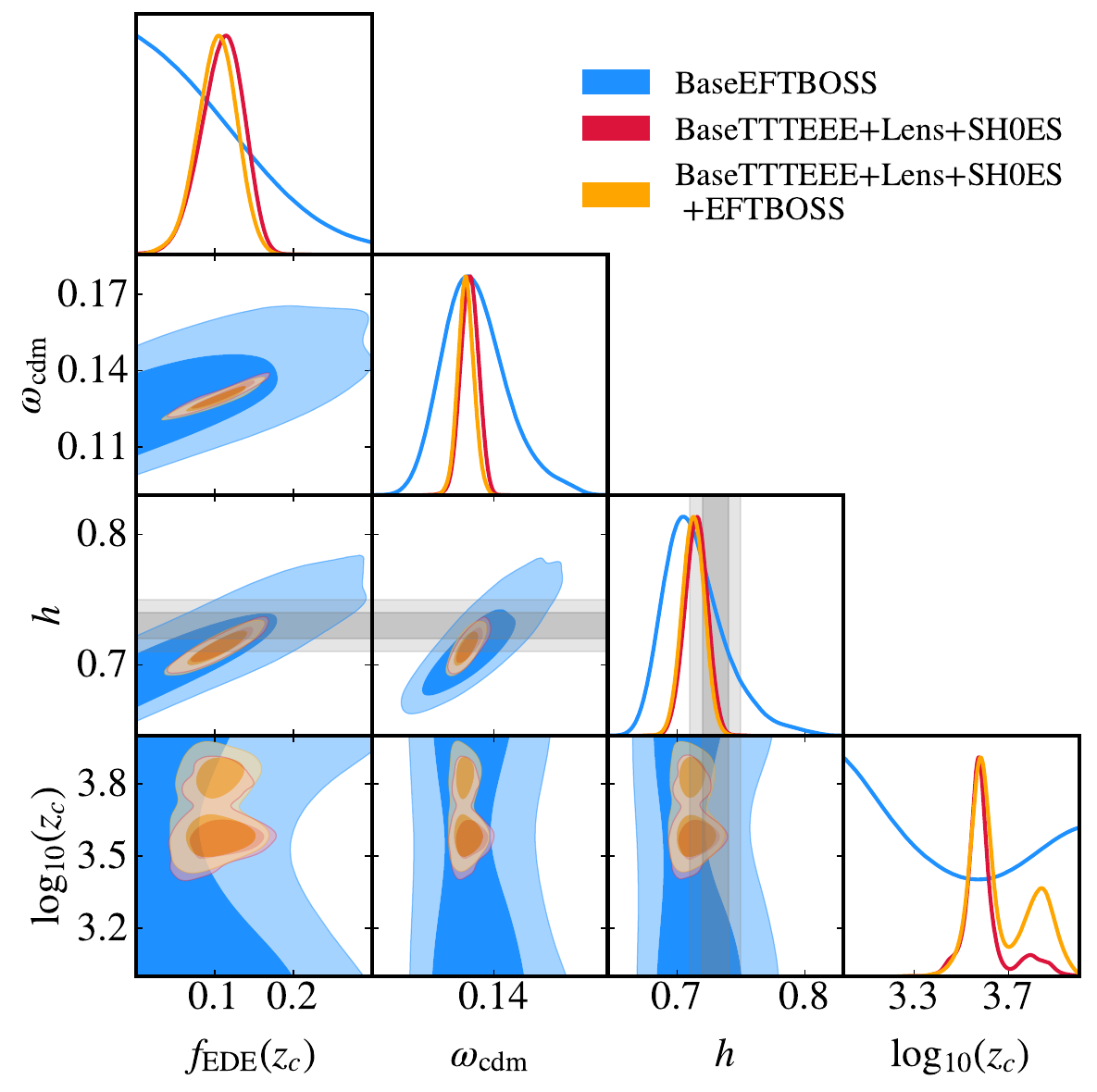}
    
    \caption{{\em Left panel:} 2D posterior distributions from BaseTTTEEE+Lens, BaseTTTEEE+Lens+$f\sigma_8$, and BaseTTTEEE+Lens+EFTBOSS. We also show the results from the EFTBOSS data with a wrong normalization for comparison. {\em Right panel:} 2D posterior distributions from BaseEFTBOSS and BaseTTTTEEE+Lens+SH0ES, with and without EFTBOSS data. We recall that BaseTTTTEEE refers to \Planck TTTEE+BAO+Pan18, while BaseEFTBOSS refers to EFTBOSS+BBN+Lens+BAO+Pan18.  }
    \label{fig:EDE_EFT_Planck}
\end{figure*}

To fully gauge the constraining power of a primary CMB-free analysis, on top of the fiducial EFTBOSS data and BBN prior, we now include other BOSS BAO measurements, {\em Planck} lensing and the Pantheon18 datasets. We recall that this dataset is simply called BaseEFTBOSS, and
we plot the associated reconstructed 2D posteriors in Fig.~\ref{fig:EDE_EFT} (blue contours). We compare our results with the posteriors reconstructed from a  BaseTTTEEE+Lens+SH0ES (red contours) and BaseTT650TEEE+ACT (orange contours) analysis. 
One can see that, while the primary CMB-free analysis does not favor EDE (in the absence of a SH0ES prior), constraints are relatively weak and the reconstructed posteriors from the BaseEFTBOSS data are not in tension with those reconstructed from the BaseTTTEEE+Lens+SH0ES and BaseTT650TEEE+ACT analyses.
Nevertheless, we note a clear narrowing of the constraints in the $\{f_{\rm EDE}(z_c),\log_{10}(z_c)\}$ parameter space around $\log_{10}(z_c)\sim 3.5$, indicating that BOSS gains constraining power right around matter-radiation equality. 
To extract a meaningful CMB-independent bound on $f_{\rm EDE}(z_c)$, we  perform an additional analysis now restricting the $\log_{10}(z_c)$ range to $\log_{10}(z_c)\in[3.4,3.7]$, which corresponds to the region favored to resolve the Hubble tension. 
We find that the combination of EFTBOSS+BBN+Lens+BAO+Pan18 (\emph{i.e.}, BaseEFTBOSS) leads to $f_{\rm EDE}(z_c)<0.2$ (95\% C.L.) and $h = 0.710_{-0.025}^{+0.015}$, which does not exclude the EDE models resolving the Hubble tension. 
When performing the same analysis with \code{CLASS-PT}, we find significantly weaker constraints, with $f_{\rm EDE}(z_c)<0.284$ (95\% C.L.) and $h =0.726_{-0.04}^{+0.02}$.
Constraints from \code{CLASS-PT} are shown in App.~\ref{app:classpt_vs_pybird_EDE}, Fig.~\ref{fig:EDE_PyBird_vs_CLASSPT}.

\section{EFTBOSS combined with CMB data}
\label{sec:EFTCMB}

\begin{table*}[]
    \centering
    \scalebox{0.8}{
    \begin{tabular}{|l|c|c|c|c|c|c|}
    
    \hline
      & \multicolumn{2}{|c|}{BaseTTTEEE+Lens}  & \multicolumn{2}{|c|}{BaseTTTEEE+Lens+$f\sigma_8$}&   \multicolumn{2}{|c|}{BaseTTTEEE+Lens+EFTBOSS}\\
              \hline
              \hline

     $H_0$ prior? & no & yes & no & yes & no & yes\\
        \hline

$f_{\rm EDE}(z_c)$
	 & $< 0.091(0.088)$ 
	 & $0.109(0.122)^{+0.030}_{-0.024}$ 
	 & $< 0.088(0.057)$ 
	 & $0.102(0.118)^{+0.030}_{-0.024}$ 
	 & $< 0.083(0.082)$ 
	 & $0.103(0.116)^{+0.027}_{-0.023}$ 
	 \\
$\log_{10}(z_c)$
	 & unconstrained $(3.55)$ 
	 & $3.599(3.568)^{+0.029}_{-0.081}$ 
	 & unconstrained (3.78) 
	 & $3.603(3.569)^{+0.037}_{-0.11}$ 
	 & unconstrained (3.82)
	 & $3.67(3.83)^{+0.21}_{-0.15}$ 
	 \\
$\theta_i$
	 & unconstrained (2.8)
	 & $2.65(2.73)^{+0.22}_{-0.025}$ 
	 & unconstrained (2.94) 
	 & $2.58(2.76)^{+0.33}_{+0.034}$ 
	 & unconstrained (2.9)
	 & $2.73(2.89)^{+0.19}_{-0.065}$ 
	 \\
	 \hline
	 $h$
	 & $0.688(0.706)^{+0.006}_{-0.011}$ 
	 & $0.715(0.719)\pm 0.009$ 
	 & $0.687(0.694)^{+0.006}_{-0.011}$ 
	 & $0.712(0.718)\pm 0.009$ 
	 & $0.687(0.700)^{+0.006}_{-0.011}$ 
	 & $0.713(0.715)\pm 0.009$ 
	 \\
$\omega_{\rm cdm }$
	 & $0.1227(0.1281)^{+0.0018}_{-0.0036}$ 
	 & $0.1303(0.1319)\pm 0.0035$ 
	 & $0.1227(0.1246)^{+0.0016}_{-0.0036}$ 
	 & $0.1296(0.1314)\pm 0.0035$ 
	 & $0.1221(0.1269)^{+0.0015}_{-0.0033}$ 
	 & $0.1288(0.1297)\pm 0.0032$ 
	 \\
$10^{2}\omega_{b }$
	 & $2.258(2.266)^{+0.018}_{-0.020}$ 
	 & $2.283(2.303)\pm 0.020$ 
	 & $2.258(2.266)^{+0.017}_{-0.021}$ 
	 & $2.282(2.279)\pm 0.021$ 
	 & $2.257(2.275)^{+0.017}_{-0.020}$ 
	 & $2.287(2.301)\pm 0.023$ 
	 \\
$10^{9}A_{s }$
	 & $2.122(2.135)\pm 0.032$ 
	 & $2.153(2.145)\pm 0.032$ 
	 & $2.119(2.119)^{+0.029}_{-0.033}$ 
	 & $2.146(2.164)\pm 0.031$ 
	 & $2.113(2.120)\pm 0.032$ 
	 & $2.144(2.144)\pm 0.032$ 
	 \\
$n_{s }$
	 & $0.9734(0.9823)^{+0.0053}_{-0.0076}$ 
	 & $0.9883(0.9895)\pm 0.0060$ 
	 & $0.9730(0.9809)^{+0.0048}_{-0.0074}$ 
	 & $0.9868(0.9899)\pm 0.0062$ 
	 & $0.9715(0.9827)^{+0.0049}_{-0.0076}$ 
	 & $0.9867(0.9921)\pm 0.0065$ 
	 \\
$\tau_{\rm reio }$
	 & $0.0570(0.0574)^{+0.0069}_{-0.0076}$ 
	 & $0.0582(0.0579)\pm 0.0075$ 
	 & $0.0564(0.0553)\pm 0.0072$ 
	 & $0.0572(0.059)\pm 0.0073$ 
	 & $0.0562(0.0553)\pm 0.0073$ 
	 & $0.0586(0.0599)^{+0.0068}_{-0.0076}$ 
	 \\
	 \hline
	 $S_8$
	 & $0.831(0.839)^{+0.011}_{-0.013}$ 
	 & $0.839(0.843)\pm 0.012$ 
	 & $0.831(0.833)^{+0.011}_{-0.012}$ 
	 & $0.838(0.843)\pm 0.013$ 
	 & $0.826(0.836)\pm 0.011$ 
	 & $0.833(0.835)\pm 0.012$ 
	 \\
$\Omega_{m }$
	 & $0.3084(0.3041)\pm 0.0058$ 
	 & $0.3008(0.3005)\pm 0.0048$ 
	 & $0.3089(0.3074)\pm 0.0054$ 
	 & $0.3019(0.3003)\pm 0.0051$ 
	 & $0.3077(0.3065)\pm 0.0054$ 
	 & $0.2998(0.3004)\pm 0.0050$ 
	 \\

   \hline
    total $\chi^2_{\rm min}$ & 3799.2&3802.9  & 3801.8& 3806.1 & 3912.7 & 3917.3\\
    $\Delta \chi^2_{\rm min}$ & -3.8 & -23.7 & -3.9&-23.0  & -4.7 & -22.7\\
    
    \hline
    $Q_{\rm DMAP}$&\multicolumn{2}{|c|}{1.9$\sigma$} &\multicolumn{2}{|c|}{2.0$\sigma$}& \multicolumn{2}{|c|}{2.1$\sigma$}\\
    \hline

    \end{tabular}}
    \caption{ Mean (best-fit) $\pm 1\sigma$ (or $2\sigma$ for one-sided bounds) of reconstructed parameters in the EDE model confronted to various datasets, including \Planck TTTEEE.}
    \label{tab:cosmoparam_planck}
\end{table*}

\begin{table*}[]
    \centering
    \scalebox{0.9}{
    \begin{tabular}{|l|c|c|c|c|}

        \hline
        & BaseTT650TEEE+ACT &  BaseTT650TEEE+ACT & BaseTT650TEEE+ACT &  BaseTT650TEEE+ACT \\
     &  &+$f\sigma_8$ &  +EFTBOSS & +Lens+EFTBOSS\\
        \hline
        \hline

$f_{\rm EDE}(z_c)$
	 & $0.128(0.159)^{+0.064}_{-0.039}$ 
	 & $0.116(0.148)^{+0.059}_{-0.046}$ 
	 & $0.093(0.148)^{+0.047}_{-0.066}$ 
	 & $< 0.172(0.148)$ 
	 \\
$\log_{10}(z_c)$
	 & $3.509(3.521)^{+0.048}_{-0.033}$ 
	 & $3.505(3.514)^{+0.056}_{-0.049}$ 
	 & $3.493(3.514)^{+0.080}_{-0.093}$ 
	 & $3.486(3.514)^{+0.091}_{-0.13}$ 
	 \\
	 
$\theta_i$
	 & $2.63(2.77)^{+0.24}_{+0.023}$ 
	 & $2.53(2.78)^{+0.37}_{+0.094}$ 
	 & $2.54(2.78)_{0.065}^{+0.47}$
	 & $2.41(2.78)_{0.12}^{+0.65}$
	 \\
	 
\hline
$h$
	 & $0.723(0.733)^{+0.021}_{-0.017}$ 
	 & $0.718(0.728)\pm 0.018$ 
	 & $0.713( 0.730)^{+0.017}_{-0.021}$ 
	 & $0.708(0.725)^{+0.015}_{-0.022}$ 
	 \\	 
$\omega{}_{\rm cdm }$
	 & $0.1332(0.1369)^{+0.0071}_{-0.0059}$ 
	 & $0.1320(0.1355)\pm 0.0062$ 
	 & $0.1285(0.1355)^{+0.0057}_{-0.0067}$ 
	 & $0.1276(0.1355)^{+0.0047}_{-0.0074}$ 
	 \\
$10^{2}\omega{}_{b }$
	 & $2.268( 2.267)\pm 0.019$ 
	 & $2.266(2.261)\pm 0.020$ 
	 & $2.265(2.266)\pm 0.020$ 
	 & $2.263(2.265)\pm 0.019$ 
	 \\
$10^{9}A_{s }$
	 & $2.144(2.148)\pm 0.037$ 
	 & $2.136(2.144)\pm 0.038$ 
	 & $2.128(2.147)\pm 0.040$ 
	 & $2.127(2.143)\pm 0.034$ 
	 \\
$n_{s }$
	 & $0.9928(0.9963)^{+0.0092}_{-0.0078}$ 
	 & $0.9910(0.9936)^{+0.0090}_{-0.0081}$ 
	 & $0.9885(0.9936)\pm 0.0091$ 
	 & $0.9865(0.9936)\pm 0.0086$ 
	 \\
$\tau{}_{\rm reio }$
	 & $0.0520(0.0508)\pm 0.0077$ 
	 & $0.0511(0.0506)\pm 0.0079$ 
	 & $0.0519(0.0506)\pm 0.0077$ 
	 & $0.0523(0.0506)\pm 0.0072$ 
	 \\
	 \hline
$S_8$
	 & $0.842(0.846)\pm 0.016$ 
	 & $0.841(0.845)\pm 0.017$ 
	 & $0.830(0.838)\pm 0.016$ 
	 & $0.831(0.837)^{+0.013}_{-0.014}$ 
	 \\	 
$\Omega{}_{m }$
	 & $0.2996(0.2982)^{+0.0061}_{-0.0072}$ 
	 & $0.3013(0.2995)\pm 0.0068$ 
	 & $0.2990(0.2995)\pm 0.0069$ 
	 & $0.3008(0.2995)\pm 0.0059$ 
	 \\

      \hline
      total $\chi^2_{\rm min}$ & 3571.9&3575.8 &3688.3 &3698.4 \\
            $\Delta\chi^2$(EDE$-\Lambda$CDM)& -14.6 & -13.3 & -12.0 & -11.1 \\
        \hline
    \end{tabular}}
    \caption{ Mean (best-fit) $\pm 1\sigma$ (or $2\sigma$ for one-sided bounds) of reconstructed parameters in the EDE model confronted to various datasets, including \Planck TT650TEEE+ACT.}
    \label{tab:cosmoparam_act}
\end{table*}

\subsection{EFTBOSS+{\em Planck}TTTEEE}
We now turn to studying the constraining power of EFTBOSS data in combination with primary CMB datasets. 
We start by performing joint analyses with the full {\em Planck}TTTEEE datasets. 
All relevant $\chi^2$ statistics are given in App.~\ref{app:chi2}, Tabs.~\ref{tab:chi2_Planck_LCDM} and \ref{tab:chi2_Planck_EDE}, while the reconstructed posteriors and best-fit values of parameters are given in Tab.~\ref{tab:cosmoparam_planck}. In the left panel of Fig.~\ref{fig:EDE_EFT_Planck}, we compare constraints obtained with the consistently and inconsistently normalized EFTBOSS data to that obtained with the compressed BAO/$f\sigma_8$ data.  
One can see that the correction of the normalization of the window function leads the new EFTBOSS data to have a constraining power only slightly stronger than the compressed BAO/$f\sigma_8$ data. We derive a BaseTTTEEE+Lens+EFTBOSS constraint of $f_{\rm EDE}(z_c)<0.083$, to be compared with  $f_{\rm EDE}(z_c)<0.088$ from BaseTTTEEE+Lens+$f\sigma_8$, while the EFTBOSS data with wrong normalization incorrectly lead to $f_{\rm EDE}(z_c)<0.054$.

Moreover, as was already pointed out in various works \cite{Murgia:2020ryi,Smith:2020rxx,Schoneberg:2021qvd,Herold:2021ksg}, posteriors are highly non-Gaussian with long tails toward high-$H_0$, and therefore these constraints should be interpreted with care. This is further attested by the fact that the best-fit point lies at the $2\sigma$ limit of our constraints (\emph{e.g.},~$f_{\rm EDE}$ at the best-fit is $0.082$ for BaseTTTEEE+Lens+EFTBOSS). We defer to future work to compare constraints derived here with a Bayesian analysis to those derived with a profile likelihood approach (\emph{e.g.}, \cite{Herold:2021ksg,Reeves:2022aoi}), which will be affected by our update to the survey window function calculation.

As advocated recently, we will gauge the level of the Hubble tension by computing the tension metric $Q_{\rm DMAP}\equiv \sqrt{\chi^2({\rm w/~SH0ES})-\chi^2({\rm w/o~SH0ES})}$ \cite{Raveri:2018wln,Schoneberg:2021qvd}, which agrees with the usual Gaussian metric tension for Gaussian posteriors, but better captures the non-Gaussianity of the posterior. 

Once the SH0ES prior is included in the BaseTTTEEE+Lens+EFTBOSS analysis, we reconstruct $f_{\rm EDE}(z_c)=0.103^{+0.027}_{-0.023}$ with $h = 0.713\pm0.009$ and find the tension metric $Q_{\rm DMAP}=2.1\sigma$ (while we find 4.8$\sigma$ in $\Lambda$CDM),  see Tab.~\ref{tab:cosmoparam_planck} and  Fig.~\ref{fig:EDE_EFT_Planck}, right panel. 
This is only a minor difference compared to the results without BOSS $f\sigma_8$ or full-shape information, for which we get $f_{\rm EDE}(z_c)=0.109^{+0.030}_{-0.024}$ with $h = 0.715\pm0.009$ and the $Q_{\rm DMAP}$ metric gives a $1.9\sigma$ tension between SH0ES and other datasets.\footnote{This is different than what was reported in Ref.~\cite{Schoneberg:2021qvd}, because of an updated $H_0$ prior with tighter error bars.} Similarly, when the $f\sigma_8$ information is included, we find  a $2.0\sigma$ tension with $f_{\rm EDE}(z_c)=0.102^{+0.030}_{-0.024}$ and $h = 0.712\pm0.009$.

Analyses with \code{CLASS-PT} are presented in App.~\ref{app:classpt_vs_pybird_EDE}, and similar results are found. Therefore, current full-shape EFTBOSS data provide little additional constraining power ($\sim 10\%$) on the EDE model over {\em Planck} and $f\sigma_8$. 
We conclude that the EFTBOSS data are in agreement with the model reconstructed when including a SH0ES prior, as the preliminary study suggested, and BOSS data do not exclude the EDE resolution to the Hubble tension.

\begin{figure*}
    \centering

\includegraphics[width=1.\columnwidth]{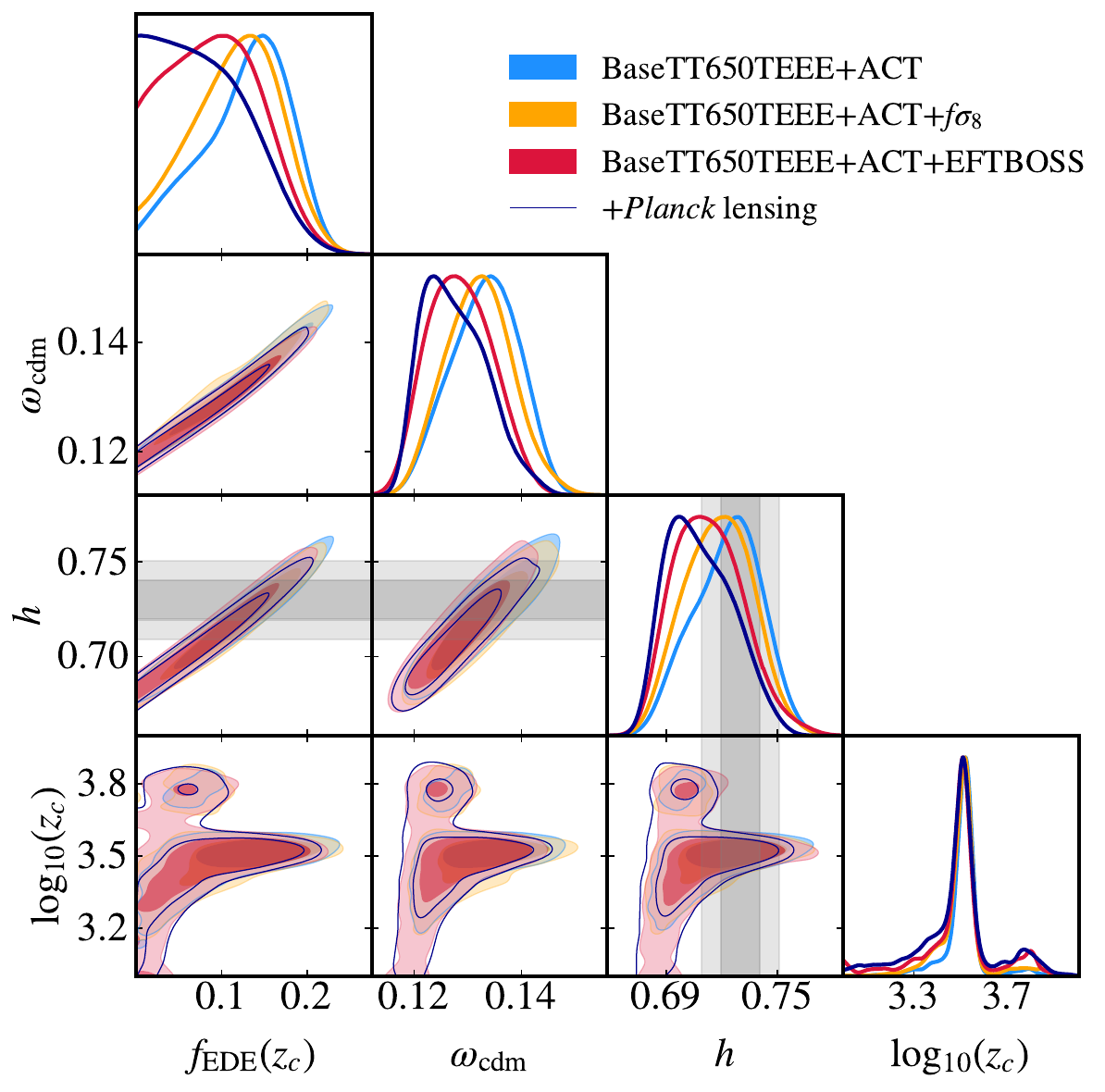}
\includegraphics[width=1.\columnwidth]{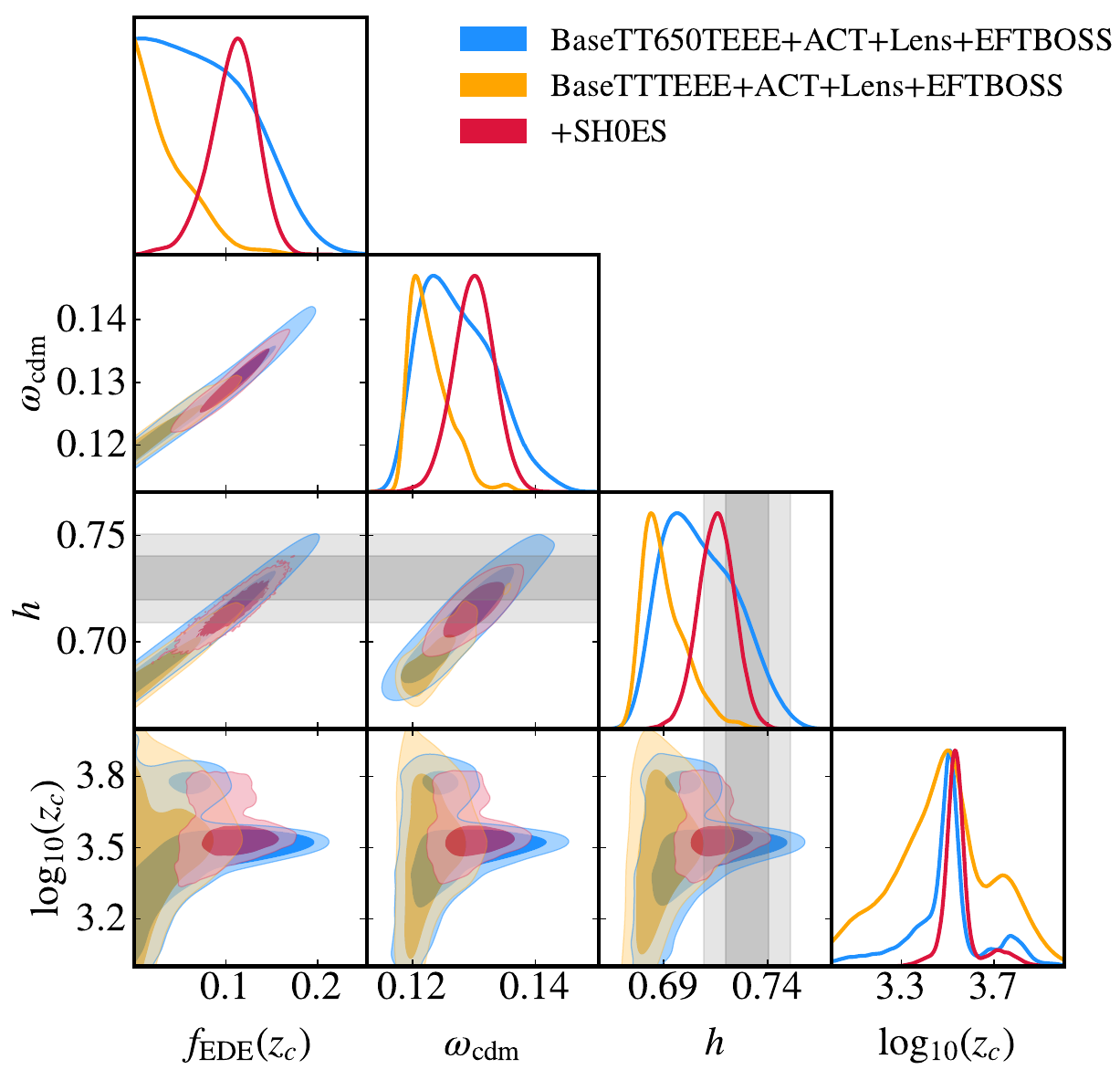}

\caption{{\em Left Panel:} 2D posterior distributions from BaseTT650TEEE+ACT in combination with $f\sigma_8$, EFTBOSS, and \Planck{} lensing. We recall that BaseTT650TEEE refers to \Planck TT650TEEE+BAO+Pan18 data. 
{\em Right Panel:}  2D posterior distributions from ACT+Lens+EFTBOSS in combination with either BaseTT650TEEE or BaseTTTEEEE  with and without SH0ES. }

    \label{fig:EDE_EFT_ACT}
\end{figure*}

\subsection{EFTBOSS+{\em Planck}TT650TEE+ACT}

We now turn to the combination of {\em Planck} data with ACT. We start with a restricted version of {\em Planck} temperature data at $\ell<650$ (chosen to mimic WMAP and perform a consistency test between CMB datasets), combined with {\em Planck} polarization and ACT data. This data combination\footnote{The preference persists until {\em Planck}TT data at $\ell\gtrsim1300$ are included, while the inclusion of SPT-3G TEEE data has little impact (in fact, slightly strengthening the hint of EDE) \cite{Smith:2022hwi}. } is known to favor\footnote{As discussed by the ACT Collaboration \cite{Hill:2021yec}, it is still a possibility that the apparent preference for EDE arises from remaining systematic errors in the data.} EDE at $\sim 3\sigma$ \cite{Hill:2021yec,Poulin:2021bjr,LaPosta:2021pgm,Smith:2022hwi}, with large values of $f_{\rm EDE}(z_c)=0.128^{+0.064}_{-0.039}$ and $h =0.723^{+0.021}_{-0.017} $ (see Tab.~\ref{tab:cosmoparam_act}, first column). In Ref.~\cite{Smith:2022hwi}, it was shown that BOSS $f\sigma_8$ and {\em Planck} lensing data decreased the preference\footnote{In the following, the preference is computed assuming the $\Delta\chi^2$ follows a $\chi^2$ distribution with three degrees of freedom. We stress that this is just an approximation, as the true number of degrees of freedom is more complicated to estimate due to $\log_{10}(z_c)$ and $\theta_i$ becoming ill defined when $f_{\rm EDE}\to 0$.} to 2.6$\sigma$.
We now test whether the EFT analysis of BOSS data can put further pressure on this hint of EDE, as our preliminary study indicates.  All relevant $\chi^2$ statistics are given in App.~\ref{app:chi2}, Tab.~\ref{tab:chi2_ACT}, while we give the reconstructed posteriors of parameters in Tab.~\ref{tab:cosmoparam_act}.
We show in Fig.~\ref{fig:EDE_EFT_ACT} (left panel) the 2D posterior distribution $\{f_{\rm EDE}(z_c),\omega_{\rm cdm},h,  \log_{10}(z_c) \}$ reconstructed from the analysis of BaseTT650TEEE+ACT compared with that reconstructed with the addition of either $f\sigma_8$ or EFTBOSS data.

One can see that, in this case, the EFTBOSS data do reduce the preference for EDE, with $f_{\rm EDE}$ now compatible with zero at $1\sigma$. For the BaseTT650TEEE +ACT+Lens+EFTBOSS dataset,  represented by the dark blue line on Fig.~\ref{fig:EDE_EFT_ACT} (left panel), we find a weak upper limit $f_{\rm EDE} < 0.172$ and $h = 0.708^{+0.015}_{-0.022}$,  with  best-fit values  $f_{\rm EDE}\simeq 0.148$ and $h\simeq 0.725$ in good agreement with the SH0ES determination. 
Quantifying the preference over $\Lambda$CDM, we find a  $\Delta\chi^2 = -11.1$ in favor of EDE (2.5$\sigma$), decreased from $-14.6$ without EFTBOSS and {\em Planck} lensing data. 
The $\chi^2$ of EFTBOSS data is degraded by $+1.7$ in the EDE model compared to $\Lambda$CDM, while the improvement in the fit of ACT and \Planck TT650TEEE is fairly stable, with $\Delta\chi^2({\rm ACT})=-7.6$ and $\Delta \chi^2({\rm \Planck TT650TEEE}) = -6.1$, respectively.
Additionally, we note that, for this more extreme EDE model, the full EFTBOSS data provide stronger constraints than the conventional BAO/$f\sigma_8$ data.
Although current data do not fully erase the preference for EDE over $\Lambda$CDM, this confirms that BOSS data, and more generally measurement of the matter power spectrum in the late Universe, provide an important probe of large EDE fraction in the early Universe. 
We find similar results with \code{CLASS-PT} (see App.~\ref{app:classpt_vs_pybird_EDE} for details), attesting that once BOSS data are combined with CMB data, the results obtained are robust to reasonable choices in the EFT analysis.

\begin{table*}[]
    \centering
    \scalebox{1}{
    \begin{tabular}{|l|c|c|}

    \hline

 & \multicolumn{2}{|c|}{BaseTTTEEE+ACT+Lens+EFTBOSS}\\ 
              \hline
    \hline
     $H_0$ prior? & no & yes \\

\hline
$f_{\rm EDE}(z_c)$
	 & $< 0.110 (0.074)$ 
	 & $0.108(0.124)^{+0.028}_{-0.021}$ 
	 \\
$\log_{10}(z_c)$
	 & $3.48(3.51)\pm 0.21$ 
	 & $3.552(3.531)^{+0.026}_{-0.065}$ 
	 \\
$\theta_i$
	 & unconstrained 
	 & $2.77(2.81)^{+0.13}_{-0.070}$ 
	 \\
	 \hline
	 $h$
	 & $0.691(0.7)^{+0.006}_{-0.013}$ 
	 & $0.715(0.72)\pm 0.009$ 
	 \\
$\omega{}_{\rm cdm }$
	 & $0.1229(0.1267)^{+0.0017}_{-0.0042}$ 
	 & $0.1300(0.1322)^{+0.0035}_{-0.0031}$ 
	 \\
$10^{2}\omega{}_{b }$
	 & $2.247(2.248)^{+0.015}_{-0.017}$ 
	 & $2.260(2.255)\pm 0.018$ 
	 \\
$10^{9}A_{s }$
	 & $2.126(2.133)^{+0.028}_{-0.032}$ 
	 & $2.153(2.156)\pm 0.030$ 
	 \\
$n_{s }$
	 & $0.9758(0.9795)^{+0.0049}_{-0.0080}$ 
	 & $0.9873(0.9893)\pm 0.0058$ 
	 \\
$\tau{}_{\rm reio }$
	 & $0.0540(0.0534)\pm 0.0070$ 
	 & $0.0548(0.0539)\pm 0.0070$ 
	 \\
	 \hline
	    $S_8$
	 & $0.829(0.843)^{+0.010}_{-0.012}$ 
	 & $0.837(0.843)\pm 0.012$ 
	 \\
$\Omega{}_{m }$
	 & $0.3061(0.3052)\pm 0.0054$ 
	 & $0.2997(0.3)\pm 0.0047$ 
	 \\
\hline
    total $\chi^2_{\rm min}$ & 4157.6 & 4159.8  \\
    $\Delta \chi^2_{\rm min}({\rm EDE}-\Lambda{\rm CDM})$  &-6.4 &-26.1 \\
  
       \hline
    $Q_{\rm DMAP}$ &\multicolumn{2}{|c|}{1.5$\sigma$}\\
\hline 
    \end{tabular}}
    \caption{ Mean (best-fit) $\pm 1\sigma$ (or $2\sigma$ for one-sided bounds) of reconstructed parameters in the EDE model confronted to BaseTTTEEE+ACT+Lens+EFTBOSS, with and without SH0ES.}
    \label{tab:cosmoparam_planck_act}
\end{table*}
\subsection{EFTBOSS+{\em Planck}TTTEE+ACT}

Except for consistency tests, there are no good reasons to remove part of the high-$\ell$ \Planck{} TT data.  In the following, we present results of combined analyses of {\em Planck}TTTEEE+ACT+EFTBOSS (\emph{i.e.}, including full \Planck{} data) in Tab.~\ref{tab:cosmoparam_planck_act} and Fig.~\ref{fig:EDE_EFT_ACT} (right panel). All relevant $\chi^2$ statistics are given in App.~\ref{app:chi2}, Tab.~\ref{tab:chi2_FullPlanckACT}. We quantify the residual tension with SH0ES using the $Q_{\rm DMAP}$ metric introduced previously. In that case, we find that the preference for EDE without SH0ES is strongly reduced, in agreement with previous works, but the $2\sigma$ upper limit on $f_{\rm EDE} < 0.110 $ is much weaker than in the BaseTTTEEE+Lens+EFTBOSS analysis presented previously,  $f_{\rm EDE} < 0.083$.
As a result, the tension metric between BaseTTTEEE+ACT+Lens+EFTBOSS and  SH0ES is released to $1.5\sigma$ compared to $ 4.7\sigma$ in $\Lambda$CDM (and $2.1\sigma$ without ACT data). When the SH0ES prior is included, we find $f_{\rm EDE} =0.108_{-0.021}^{+0.028}$ and $h = 0.715\pm0.009$ (in very good agreement with the results presented earlier without ACT), with no degradation in the $\chi^2$ of EFTBOSS. This confirms that the EFTBOSS data can accommodate the amount of EDE required to resolve the Hubble tension (with $f_{\rm EDE}\sim 0.1$ and $h\sim 0.72$), but constrain more extreme EDE contributions. 

\subsection{Impact of Pantheon+ data}
\label{sec:PanPlus}
\begin{table*}[]
    \centering
    \scalebox{0.9}{
    \begin{tabular}{|l|c|c|c|c|}

    \hline
&   BaseEFTBOSS & BaseTTTEEE+Lens & BaseTTTEEE+Lens& BaseTT650TEEE+ACT+Lens\\
 & +PanPlus& +EFTBOSS+PanPlus & +EFTBOSS+PanPlus+SH0ES & +EFTBOSS+PanPlus\\
        \hline
     \hline

$f_{\rm EDE}(z_c)$
	 & $ < 0.228(0.01)$ 
	 & $< 0.079(0.056)$ 
	 & $0.123(0.141)^{+0.030}_{-0.018}$ 
	 & $ < 0.137(0.11)$ 
	 \\
$\log_{10}(z_c)$
	 & unconstrained $(3.91)$ 
	 & $3.59(3.57)^{+0.25}_{-0.21}$ 
	 & $3.64(3.57)^{+0.23}_{-0.13}$ 
	 & $< 3.5(3.5)$ 
	 \\
$\theta_i$
	 & unconstrained$(2.98)$ 
	 & unconstrained$(2.74)$ 
	 & $2.59(2.77)^{+0.31}_{+0.064}$ 
	 & unconstrained$(2.78)$ 
	 \\
  \hline
  $h$
	 & $0.717(0.692)^{+0.015}_{-0.026}$ 
	 & $0.684(0.692)^{+0.006}_{-0.001}$ 
	 & $0.719(0.724)^{+0.009}_{-0.008}$ 
	 & $0.700(0.714)^{+0.013}_{-0.019}$ 
	 \\
$\omega{}_{\rm cdm }$
	 & $0.142(0.131)^{+0.010}_{-0.014}$ 
	 & $0.1222(0.1251)^{+0.0015}_{-0.0028}$ 
	 & $0.1317(0.1346)\pm 0.0031$ 
	 & $0.1258(0.1306)^{+0.0039}_{-0.0058}$ 
	 \\
$10^{-2}\omega{}_{b }$
	 & $2.276(0.023)^{+0.035}_{-0.039}$ 
	 & $2.251(2.254)\pm 0.018$ 
	 & $2.291(2.275)^{+0.020}_{-0.024}$ 
	 & $2.258(2.259)\pm 0.019$ 
	 \\
$10^9A_s$
	 & $1.88(1.929)^{+0.16}_{-0.20}$ 
	 & $2.114(2.148)\pm 0.029$ 
	 & $2.155(2.157)^{+0.030}_{-0.036}$ 
	 & $2.120(2.135)\pm 0.033$ 
	 \\
$n_{s }$
	 & $0.873(0.889)\pm 0.049$ 
	 & $0.9700(0.9752)^{+0.0046}_{-0.0071}$ 
	 & $0.9911(0.9912)^{+0.0062}_{-0.0071}$ 
	 & $0.9827(0.9877)\pm 0.0081$ 
	 \\
$\tau_{\rm reio}$
	 & $-$	
	 & $0.0562(0.0558)\pm 0.0069$ 
	 & $0.0582(0.0554)\pm 0.0077$ 
	 & $0.0519(0.0516)^{+0.0065}_{-0.0075}$ 
	 \\
	 \hline
	  $S_8$
	 & $0.815(0.824)\pm 0.018$ 
	 & $0.832(0.837)\pm 0.010$ 
		 & $0.840(0.847)\pm 0.012$ 
	 & $0.831(0.839)^{+0.012}_{-0.011}$ 
	 \\
$\Omega{}_{m }$
	 & $0.321(0.324)\pm 0.013$ 
	 & $0.3116(0.3093)\pm 0.0056$ 
	 & $0.3000(0.3014)\pm 0.0047$ 
	 & $0.3041(0.3016)\pm 0.0061$ 
	 \\

\hline 
total $\chi^2_{\rm min}$ & 1537.9 &4304.0 & 4187.0 &4085.1 \\
    $\Delta \chi^2_{\rm min}\textrm{\scriptsize (EDE-$\Lambda$CDM)}$ & 0 &-1.1 & -32.3 & -9.2 \\
       
\hline
    \end{tabular}}
    \caption{ Mean (best-fit) $\pm 1\sigma$ (or $2\sigma$ for one-sided bounds) of reconstructed parameters in the EDE model confronted to various datasets, including the recent PanPlus SNIa catalog.}
    \label{tab:cosmoparam_PanPlu}
\end{table*}

\begin{figure*}
    \centering
\includegraphics[width=1.\columnwidth]{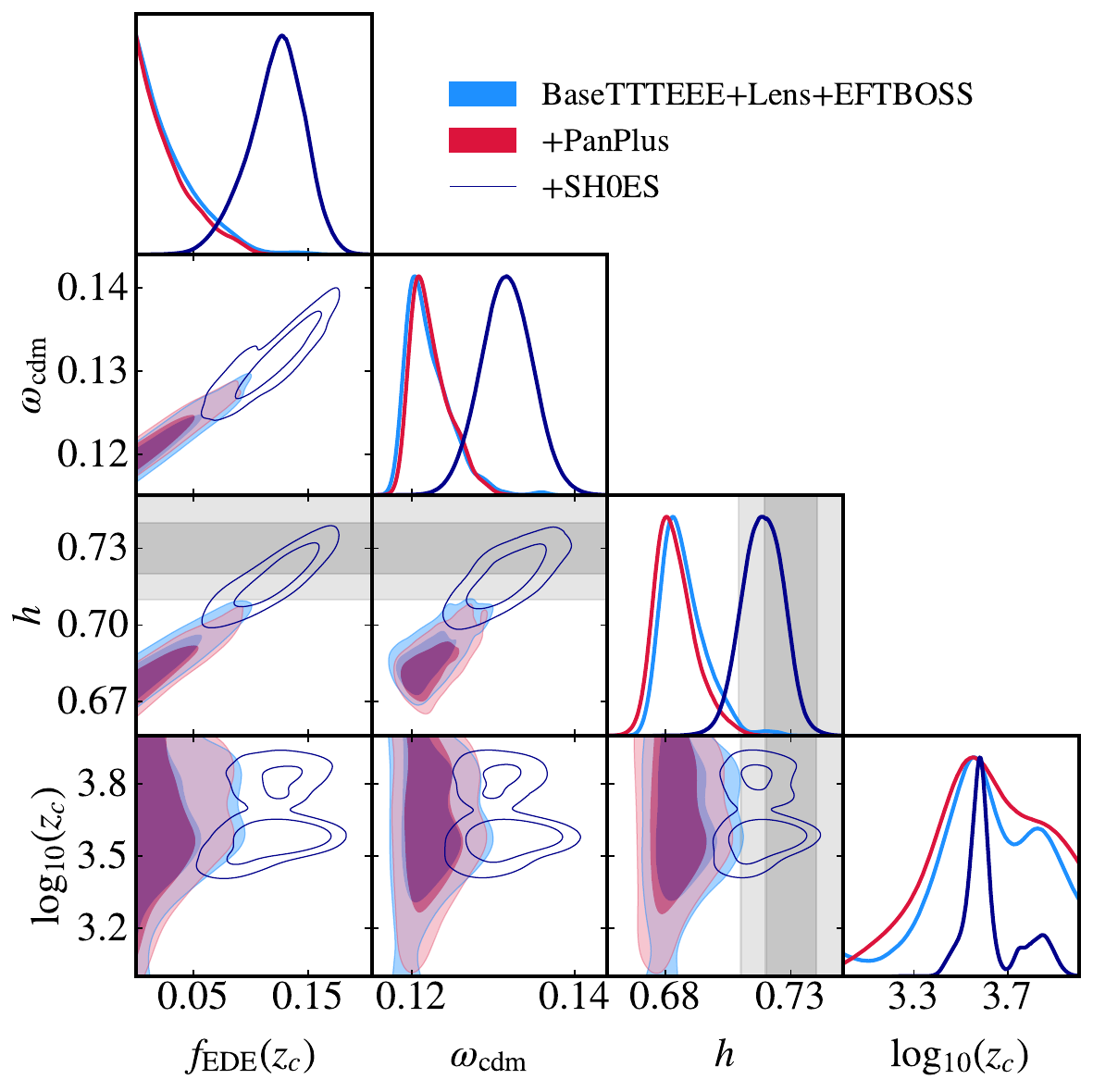}
\includegraphics[width=1.\columnwidth]{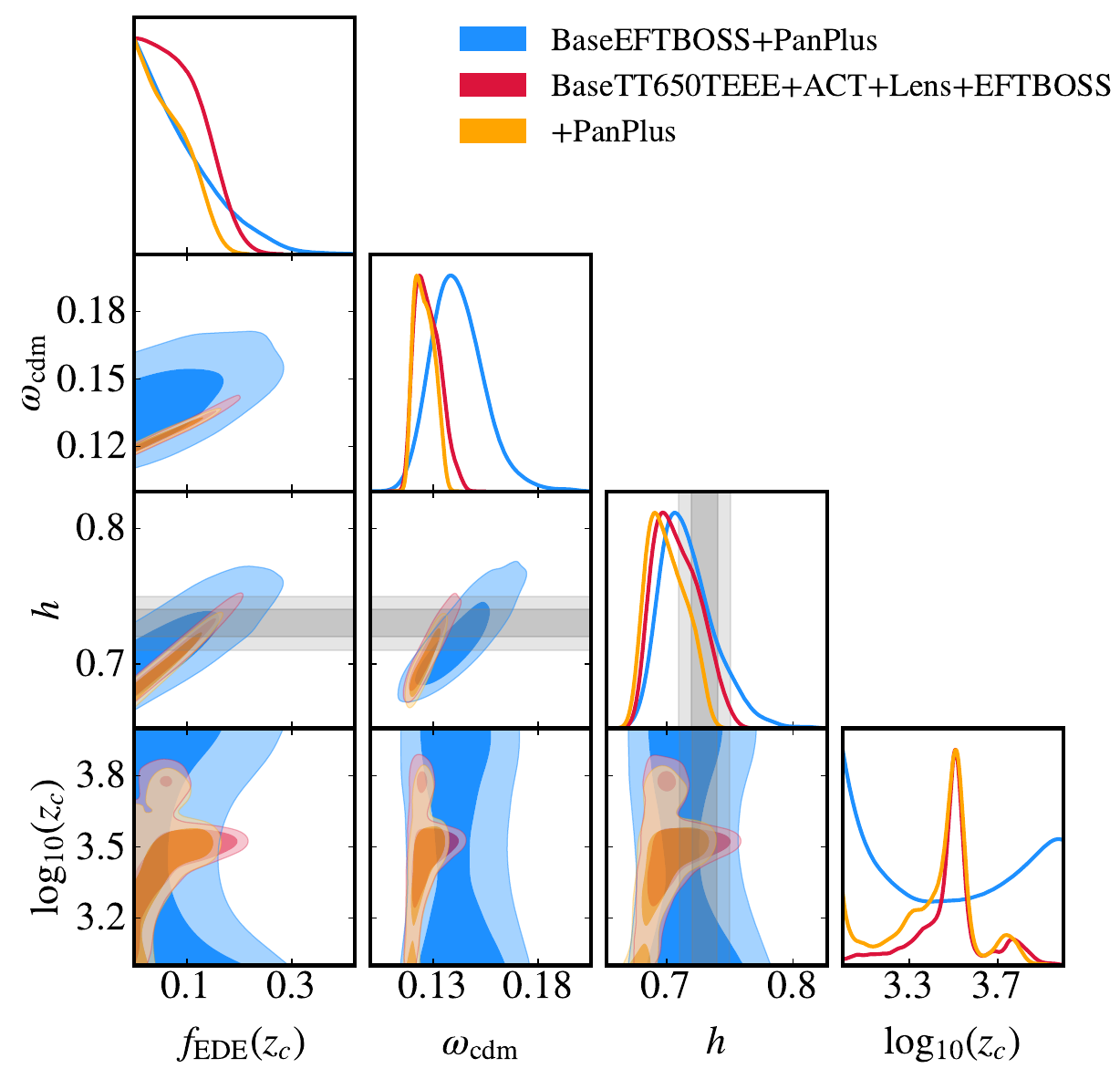}
\includegraphics[width=1.\columnwidth]{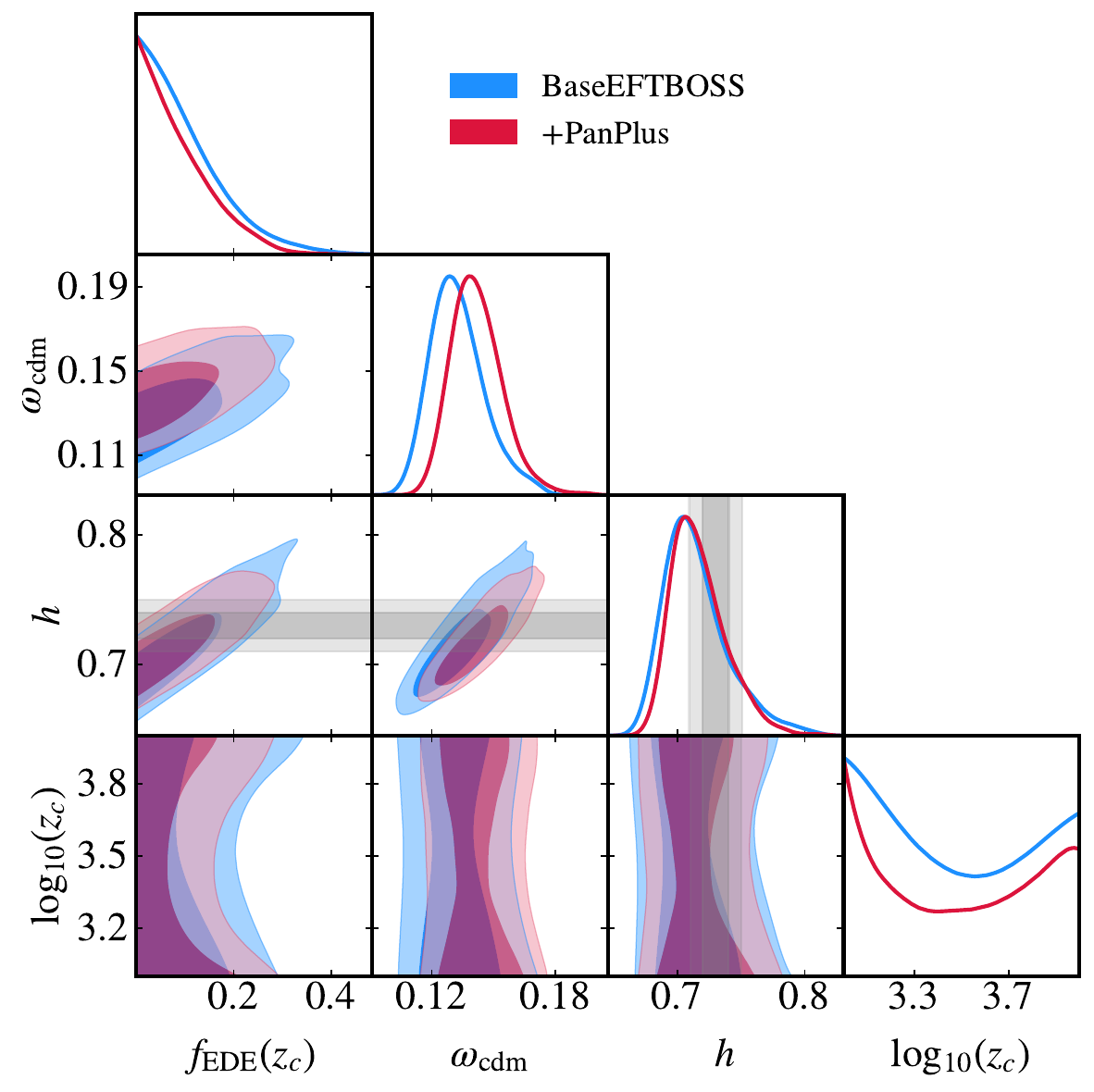}

\caption{{\em Top left panel:} 2D posterior distributions from BaseTTTEEE+Lens+EFTBOSS in combination with either Pantheon18 or Pantheon+ data, and the SH0ES cepheid calibration. We recall that BaseTTTEEE refers to \Planck TTTEEE+BAO+Pan18 data.
{\em Top right panel:} 2D posterior distributions from BaseEFTBOSS and BaseTT650TEEE+ACT+Lens+EFTBOSS, in combination with either Pantheon18 or Pantheon+ data. We recall that BaseTT650TEEE refers to \Planck TT650TEEE+BAO+Pan18 data, while BaseEFTBOSS refers to EFTBOSS+BBN+Lens+BAO+Pan18. 
{\em Bottom panel:} 2D posterior distributions from BaseEFTBOSS, with either Pantheon18 or Pantheon+ data.  }

    \label{fig:EDE_PanPlus}
\end{figure*}

To finish, we perform an analysis with the new Pantheon+ SNIa catalog \cite{Brout:2022vxf}, which is known to favor a higher $\Omega_m = 0.338\pm0.018$, to illustrate the impact that these new data have on the EDE model. 
We perform analyses of four datasets in combination with Pantheon+, following our baseline data, namely, BaseEFTBOSS,
BaseTTTEEE+Lens+EFTBOSS(+SH0ES), and BaseTT650TEEE+ACT+Lens+EFTBOSS. The results of these analyses are presented in Tab.~\ref{tab:cosmoparam_PanPlu} and in Fig.~\ref{fig:EDE_PanPlus}, while all relevant $\chi^2$ statistics are given in App.~\ref{app:chi2}, Tab.~\ref{tab:chi2_PanPlus}. 
First, without information from the primary CMB, we find that the combination of  EFTBOSS+BBN+Lens+BAO+PanPlus (\emph{i.e.}, BaseEFTBOSS+PanPlus) leads to a weak constraint on $f_{\rm EDE}(z_c)<0.228$ with $h=0.717^{+0.015}_{-0.026}$ in good agreement with SH0ES. In fact, even within $\Lambda$CDM we find $h=0.694_{-0.014}^{+0.012}$, which is not in significant tension with SH0ES. 
This data combination was recently argued to constrain new physics solution to the Hubble tension that affects the sound horizon, due to the fact that measurement of $h$ based on the scale of matter-radiation equality $k_{\rm eq}$ (which can be extracted by marginalizing over the sound horizon information\footnote{More precisely, in Refs~\cite{Philcox:2020vbm,Philcox:2021kcw,Philcox:2022sgj}, the marginalization over the sound horizon information is intended as a consistency test to be performed within $\Lambda$CDM.}) is in tension with the SH0ES measurement~\cite{Philcox:2020vbm,Philcox:2021kcw,Philcox:2022sgj}.
In our analysis, we stress that we do not marginalize over the sound horizon in the EFTBOSS analysis.
We do not expect that removing part of the data through the marginalization procedure would make BOSS data appear in strong tension with SH0ES, at least in EDE. 
Rather, we expect that constraints would significantly weaken. 
We leave for future work to test whether the determination of $h$ from $k_{\rm eq}$ is robust to changes in the cosmological model. 

When combining with \Planck TTTEEE, we find that constraints on EDE are increased by $\sim 5\%$ with respect to the analogous analysis with Pantheon18, with $f_{\rm EDE}(z_c)<0.079$.
This can be understood by noting that the larger $\Omega_m$ favored by Pantheon+, coupled with the positive correlation between $f_{\rm EDE}(z_c)-h$, can lead to high $\omega_m=\Omega_mh^2$ which are constrained by CMB data. 
However, once the SH0ES cepheid calibration of SNIa is included, we find a strong preference for EDE, with $f_{\rm EDE}(z_c)=0.123^{+0.030}_{-0.018}$ (\emph{i.e.}, nonzero at more than $5\sigma$) and a $\Delta\chi^2({\rm EDE}-\Lambda{\rm CDM})=-32.3$ (compared to $-22.7$ with Pantheon18). 
The cost in $\chi^2$ for \Planck TTTEEE+Lens and EFTBOSS compared to the analysis without the SH0ES calibration is small, with $\chi^2({\rm \Planck})$ increasing by $+2.3$ and $\chi^2({\rm EFTBOSS})$ increasing by $+0.9$, which further attests to the non-Gaussianity of the posterior in the absence of the SH0ES calibration. 
The $Q_{\rm DMAP}$ tension metric introduced earlier cannot be used as easily, due to the fact that the SH0ES data are now modeled in a more involved way, making use of a correlation matrix connecting SNIa calibrators and high$-z$ SNIa \cite{Riess:2021jrx}, rather than the simple Gaussian prior on $h$. 

Finally, when combining with \Planck~TT650TEEE and ACT, we find that the preference for EDE seen within ACT data further decreases to $\Delta\chi^2=-9.2$ (2.2$\sigma$) and we derive a limit $f_{\rm EDE}(z_c)<0.137$, with $h=0.700^{+0.013}_{-0.019}$ and a $\lesssim 2\sigma$ tension with SH0ES.
We defer to future work to further test the ability of EDE (and other promising models) to resolve the Hubble tension in light of this new Pantheon+ SNIa catalog.

\section{Discussion and Conclusions}
\label{sec:conclusions}

The developments of the predictions for the galaxy clustering statistics from the EFTofLSS have made possible the study of BOSS data beyond the conventional analyses dedicated to extracting BAO and $f\sigma_8$ information. 
There has been in the recent literature a number of studies aiming at measuring the $\Lambda$CDM parameters at precision comparable with that of \Planck~CMB data (see, \emph{e.g.}, Refs.~\cite{DAmico:2019fhj,Ivanov:2019pdj,Colas:2019ret,DAmico:2020kxu,DAmico:2020tty,Chen:2021wdi,Zhang:2021yna,Philcox:2021kcw}). 
Additionally, it was shown that BOSS full-shape data, when analyzed using the one-loop predictions from the EFTofLSS (here called EFTBOSS data), can lead to strong constraints on extension to the $\Lambda$CDM model. 
In particular, the EDE model, currently one of the most promising models to resolve the Hubble tension \cite{Poulin:2018cxd,Schoneberg:2021qvd}, was shown to be severely constrained by EFTBOSS data \cite{DAmico:2020ods,Ivanov:2020ril}. 
However, it was subsequently argued that part of the constraints may come from a mismatch in the primordial power spectrum $A_s$ amplitude between EFTBOSS and \Planck~\cite{Smith:2020rxx}. 

Recently, it was found that the original EFTBOSS data used in these analyses were affected by an inconsistency between the normalization of the survey window function and the one of the data measurements, which led to a mismatch in $A_s$. A proper reanalysis of the EFTBOSS data constraints on the EDE model was lacking until now. 

In this paper, we have performed a thorough investigation of the constraints on EDE in light of the correctly normalized EFTBOSS data and estimated the shifts introduced on the reconstructed cosmological parameters and their errors between various analysis strategies. 
A similar analysis within the $\Lambda$CDM model is presented in Sec. IV of our companion paper~\cite{Simon:2022lde}.
Our results are summarized in the following.
\\
\subsection{EFTBOSS constraints on EDE alone}
We have shown in Sec.~\ref{sec:EDE_arena_main} that, regardless of the BOSS data or the likelihood we consider, the BOSS full-shape (analyzed on their own with a BBN prior) leads to reconstructed values of $H_0$ that are compatible with what
is obtained by the SH0ES Collaboration. 
Yet, the various EFTBOSS measurements, as well as the \code{PyBird} and \code{CLASS-PT} likelihoods, do not have the same constraining power on EDE: 
\begin{itemize}
    \item[\textbullet] When using the \code{PyBird} likelihood, we found $f_{\rm EDE}(z_c) < 0.321$ when analyzing $P_\textsc{fkp}^\textsc{lz/cm} + \alpha^\textsc{lz/cm}_\text{rec}$, while analyzing $P_\textsc{quad}^{z_1/z_3} + \alpha^{z_1/z_3}_\text{rec}$ yields $f_{\rm EDE}(z_c) < 0.382$, a $\sim 20\%$ difference.\\
    \item[\textbullet]When using the same BOSS data, namely, $P_\textsc{quad}^{z_1/z_3}$, we have found that the \code{PyBird} likelihood gives $f_{\rm EDE}(z_c) < 0.382$, while the \code{CLASS-PT} likelihood gives $f_{\rm EDE}(z_c) < 0.448$, \emph{i.e.}, a $\sim 15\%$ difference.\\
\item[\textbullet] Restricting our analysis to the range of critical redshift $\log_{10}(z_c)\in[3.4,3.7]$ that can resolve the Hubble tension, we have shown that the combination of EFTBOSS+BBN+Lens+BAO+Pan18, leads to the constraints  $f_{\rm EDE}(z_c)<0.2$ (95\% C.L.) and $h = 0.710_{-0.025}^{+0.015}$ , which does not exclude the EDE models resolving the Hubble tension.  \\
\item[\textbullet] The inclusion of the recent Pantheon+ data does not affect this conclusion as we find $h=0.717^{+0.015}_{-0.026}$.  We do not expect that marginalizing over the sound horizon as done in Refs.~\cite{Philcox:2020vbm,Philcox:2021kcw,Philcox:2022sgj} would alter our conclusions, as it would simply remove information from the data. This question will be thoroughly explored elsewhere.
\end{itemize}

\subsection{{\em Planck}+EFTBOSS~constraints on EDE}

In combination with \Planck~TTTEEE data, we have shown that constraints on EDE have changed due to the correction of the normalization of the window function:
\begin{itemize}
    \item[\textbullet]    The combination of \Planck TTTEEE+Lens+BAO +Pan18+EFTBOSS leads to $f_{\rm EDE}(z_c)<0.083$, which is a $\sim10\%$ improvement over the constraints without BOSS data and a $\sim 5\%$ improvement over the constraints with conventional BAO/$f\sigma_8$ data.  Yet, this is  much weaker than the constraints reported with the incorrect normalization, namely, $f_{\rm EDE}<0.054$.
    We quantify that the Hubble tension is reduced to the $2.1\sigma$ level in the EDE cosmology ($1.9\sigma$ without EFTBOSS) compared to $4.8\sigma$ in the $\Lambda$CDM model, and we find $f_{\rm EDE}(z_c)=0.103^{+0.027}_{-0.023}$ at $z_c=3970^{+255}_{-205}$ when the SH0ES prior is included. 
    \item[\textbullet] Replacing Pantheon18 by the new Pantheon+ data improves the constraints on EDE to $f_{\rm EDE}(z_c)<0.079$. Yet, the inclusion of the SH0ES cepheid calibration leads to  $f_{\rm EDE}(z_c)=0.123^{+0.030}_{-0.018}$ at $z_c=4365^{+3000}_{-1100}$, \emph{i.e.}, a nonzero $f_{\rm EDE}(z_c)$ at more than $5\sigma$ with $\Delta\chi^2({\rm EDE}-\Lambda{\rm CDM})=-32.3$. 
    The cost in $\chi^2$ for \Planck TTTEEE+Lens and EFTBOSS compared to the analysis without the SH0ES calibration is small, with $\chi^2({\rm \Planck})$ increasing by $+2.3$ and $\chi^2({\rm EFTBOSS})$ increasing by $+0.9$, which attests to the non-Gaussianity of the posterior in the absence of the SH0ES calibration. This deserves to be studied further through a profile likelihood approach \cite{Herold:2021ksg,Reeves:2022aoi}.
\end{itemize}

\subsection{ACT+EFTBOSS constraints on EDE}

Finally, we have studied the impact of EFTBOSS data on the recent hints of EDE observed within ACT DR4 data:
\begin{itemize}

    \item[\textbullet] EFTBOSS reduces the preference for EDE over $\Lambda$CDM seen when analyzing ACT DR4, alone or in combination with restricted \Planck TT data. 
    The combination of \Planck TT650TEEE+Lens +BAO+Pan18+ACT+EFTBOSS leads to a mild constraints on $f_{\rm EDE}(z_c)<0.172$ with $\Delta\chi^2({\rm EDE}-\Lambda{\rm CDM})=-11.1$,  to be compared with $f_{\rm EDE}(z_c)=0.128^{+0.064}_{-0.039}$ without EFTBOSS+Lens, with $\Delta\chi^2({\rm EDE}-\Lambda{\rm CDM})=-14.6$.
    \item[\textbullet] The inclusion of Pantheon+ data further restricts $f_{\rm EDE}(z_c)<0.137$, with $\Delta\chi^2({\rm EDE}-\Lambda{\rm CDM})=-9.2$.
    \item[\textbullet] When full \Planck{} data are included, we derived a constraint $f_{\rm EDE}(z_c)<0.110$, which is $\sim 30\%$ weaker than without ACT data. 
    When all CMB data are included in combination with EFTBOSS, the Hubble tension is reduced to $1.5\sigma$ in the EDE model, to be compared with $4.7\sigma$ in $\Lambda$CDM. The inclusion of the SH0ES prior leads to $f_{\rm EDE}(z_c)=0.108^{+0.028}_{-0.021}$ at $z_c = 3565^{+220}_{-495}$.
\end{itemize}

We conclude that EFTBOSS data do not exclude EDE as a resolution to the Hubble tension, where we consistently find $f_{\rm EDE}(z_c)\sim 10-12\%$ at $z_c\sim3500-4000$, with $h\sim 0.72$, when the cepheid calibration is included in the analyses. 
However, EFTBOSS data do constrain very high EDE fraction as seen when analyzing ACT DR4 data. 

\subsection{Final comments}

There are a number of relevant caveats to stress regarding our analyses. First, we note that the reconstructed $S_8$ values from the various analyses that favor EDE are $\sim 2.8-3.2\sigma$ higher than those coming from weak lensing measurement (and their cross-correlation with galaxy clustering) such as DES \cite{DES:2021wwk} and KiDS \cite{Heymans:2020gsg}. 
As was already pointed out in the past, this indicates that weak lensing data (and the existence of a $S_8$ tension) could be used to further restrict the existence of EDE. Nevertheless, it has been noted that solutions to the $S_8$ tension may be due to systematic effects \cite{Amon:2022ycy} or nonlinear modeling including the effect of baryons at very small scales \cite{Amon:2022azi} or to a more complete dynamics in the dark sector \cite{McDonough:2021pdg,Sabla:2022xzj}. 
In fact, models that resolve the $S_8$ tension leave the EDE resolution unaffected \cite{Allali:2021azp,Clark:2021hlo} such that, although perhaps theoretically unappealing, it is possible that solutions to the $H_0$ and $S_8$ lie in different sectors.
We leave for future work a robust study of EDE in light of the combination of EFTBOSS and weak lensing data, which will require better handling of the modeling of physical effects at scales beyond the range of validity of our EFT.
Second, it will be very important to extend this work to include the bispectrum, which was recently analyzed at the one-loop level within $\Lambda$CDM~\cite{DAmico:2022osl,DAmico:2022gki}. 
It will also be  interesting to see if the eBOSS surveys can shed light on EDE~\cite{eBOSS:2020yzd}: although the inclusion of eBOSS BAO was shown to not significantly modify the constraints on EDE (see, \emph{e.g.},~Refs.~\cite{Schoneberg:2021qvd,DAmico:2020ods}), the analysis of the full-shape of eBOSS quasars may have the potential to put stronger limits given the large size of the survey. 
Additional constraints on EDE may also arise from measurements of the age of old objects such as globular clusters of stars \cite{Bernal:2021yli,Boylan-Kolchin:2021fvy}, or the halo mass function at high$-z$ \cite{Klypin:2020tud}. Interestingly, using $N$-body simulations, Ref.~\cite{Klypin:2020tud} showed that EDE predicts 50\% more massive clusters at $z = 1$ and twice more galaxy-mass halos at $z = 4$ than \lcdm. These predictions can be tested by observations from the James Webb Space Telescope and the first publicly available data are, in part, better fit by EDE than \lcdm\ \cite{Boylan-Kolchin:2022kae}.

To close this work, we mention that we find here in agreement with previous literature, that the cosmological data including SH0ES prefer a higher value for the spectral tilt $n_s$ in the EDE model than in $\Lambda$CDM, with $n_s \sim 1$ allowed at $\lesssim 2\sigma$ depending on the combination of data considered. 
Of interest here, we see that the inclusion of EFTBOSS data does not significantly pull back $n_s$ to lower value, and when analyzed alone (with a BBN prior) also independently favors a value of $n_s$ consistent with scale independence at $\sim 1\sigma$. 
A value of $n_s$ close to that of the Harrison–Zeldovich spectrum, when put in perspective of CMB measurements of the tensor-to-scalar ratio, would dramatically change the status of the preferred inflationary models~\cite{DAmico:2021fhz} (see also~Refs.~\cite{Kallosh:2022ggf,Jiang:2022uyg,Takahashi:2021bti}).  
Therefore, if EDE is firmly detected with future cosmological data, beyond serving as resolution of the $H_0$ tension, it would also have important consequences for early Universe physics.

\begin{acknowledgements}
We thank Adam Riess for interesting discussions and comments at various stages of the project, as well as kindly sharing Pantheon+ and SH0ES data, and Dillon Brout and Dan Scolnic for their precious help with the implementation of the likelihood into \code{MontePython}. We thank Guillermo Franco Abellán and José Louis Bernal for their contribution in the early stages of this project, and Guido D'Amico, Arnaud de Mattia, and Kevin Pardede for useful discussions. 
P. Z. would like to thank the organizers of the workshop \emph{LSS2022: Recent Developments in Theoretical Large-Scale Structure - IFPU} for hospitality in Trieste during the late stage of completion of this project. 
This work has been partly supported by the CNRS-IN2P3 grant Dark21. 
The authors acknowledge the use of computational resources from the Excellence Initiative of Aix-Marseille University (A*MIDEX) of the “Investissements d’Avenir” program. These results have also been made possible thanks to LUPM's cloud computing infrastructure founded by Ocevu labex, and France-Grilles.
This project has received support from the European Union’s Horizon 2020 research and innovation program under the Marie Skodowska-Curie grant agreement No. 860881-HIDDeN. This work used the Strelka Computing Cluster, which is run by Swarthmore College. T. L. S. is supported by NSF Grant No.~2009377, NASA Grant No.~80NSSC18K0728, and the Research Corporation.

\end{acknowledgements}

\appendix

\section{Window function normalization}\label{app:normalization}

As discussed in Refs.~\cite{deMattia:2019vdg,deMattia:2020fkb,Beutler:2021eqq} (see also~\cite{Sugiyama:2018yzo}), the window function measurements, which are required to make an accurate theoretical calculation, have to be consistently normalized with the power spectrum measurements. 
The estimator for the power spectrum we are concerned with is the FKP estimator~\cite{Feldman:1993ky}, later generalized to redshift space in Refs.~\cite{Yamamoto:2002bc,Yamamoto:2005dz}. 
For fast estimation using FFTs~\cite{Bianchi:2015oia,Scoccimarro:2015bla}, the line of sight for a given galaxy pair is chosen to be in the direction of one of galaxy in the pair, $\r_1$. 
For clarity in the discussion we are going to have next, let us first gather here pieces of derivations that can be found partially in Refs.~\cite{Beutler:2018vpe,DAmico:2019fhj}. 
It is easy to see that the expectation value of the power spectrum FKP estimator reads (see, \emph{e.g.},~\cite{Zhang:2021uyp})
\begin{widetext}
\begin{equation}
\braket{ \hat P_\ell(k) } =  \frac{2\ell+1}{N_P} \int \frac{d\Omega_k}{4\pi} d^3r_1 d^3s \, e^{-i \k \cdot \s} \Theta(\r_1) \Theta(\r_1 + \s) \bar n_w(\r_1) \bar n_w(\r_1 + \s) \xi(\s, \r_1) \mathcal{L}_\ell(\hat k \cdot \hat r_1) \, ,
\end{equation}
where $\mathcal{L}_\ell$ is the Legendre polynomial of order $\ell$. 
Here $\bar n_w(\r) \equiv w(\r) \bar n(\r)$ is the weighted mean galaxy density, with weight $w(\r)$ being the FKP weights times some correction weights (usually to account for veto and instrumental/observational systematics), $\Theta(\r)$ is one if the galaxy at position $\r$ falls inside the survey, zero otherwise, and $\xi(\s, \r_1)$ is the correlation function, with $\s$ the separation between two galaxies. 
Importantly, $N_P$ is a normalization factor that is \emph{chosen by the user}, as we will precise below. 
Using the following identity:
\begin{equation}
\int \frac{d\Omega_k}{4 \pi} e^{-i \k \cdot \s}  \mathcal{L}_\ell (\hat k \cdot \hat r_1) = (-i)^\ell j_\ell(ks) \mathcal{L}_\ell (\hat s \cdot \hat r_1) \ ,
\end{equation}
where $j_\ell$ is the spherical-Bessel function of order $\ell$,
we obtain 
\begin{equation}\label{eq:mean_fkp_estimator}
\braket{ \hat P_\ell(k)} = \frac{(2\ell+1)}{N_P} (-i)^\ell \int ds \, s^2 j_\ell(ks) \int d\Omega_s \int d^3 r_1 \Theta(\r_1) \Theta(\r_1 + \s) \bar n_w(\r_1) \bar n_w(\r_1 + \s) \xi(\s, \r_1) \mathcal{L}_\ell(\mu)  \ ,
\end{equation}
where we have introduced the notation $\mu \equiv \hat s \cdot \hat r_1$. 
We now make the following approximation. 
We assume that the redshift evolution of the correlation function can be neglected within the observational bin such that $\xi(\s, \r_1) \equiv \xi(s, \mu, r_1(z)) \simeq \xi(s, \mu, z_{\rm eff}) \equiv \xi(s, \mu)$, where the latter is evaluated at the effective redshift $z_{\rm eff}$ of the survey.~\footnote{See Ref.~~\cite{Zhang:2021uyp} for a BOSS analysis that does not rely on this approximation.} 
As such, we can pull out $\xi(s, \mu)$ from the integral over $d^3 r_1$. 
We can further expand in multipoles $\xi(s, \mu) = \sum_{\ell'} \xi_{\ell'}(s) \mathcal{L}_{\ell'}(\mu)$ to pull out $\xi_{\ell'}(s)$ from the angular integrals. 
Then, using the identity
\begin{equation}
\mathcal{L}_{\ell}(\mu) \mathcal{L}_{\ell'}(\mu) = \sum_L  \left(\begin{matrix}
\ell & L & \ell'\\
0 & 0 & 0
\end{matrix}\right)^2 (2L+1) \mathcal{L}_L(\mu) \ ,
\end{equation}
where $\left(\begin{matrix}
\ell & L & \ell'\\
0 & 0 & 0
\end{matrix}\right)$ are the Wigner 3-j symbols,
we get
\begin{equation}
    \braket{ \hat P_\ell(k)} = 4\pi (2\ell+1) (-i)^\ell \sum_{\ell', L} \left(\begin{matrix}
\ell & L & \ell'\\
0 & 0 & 0
\end{matrix}\right)^2 \int ds \, s^2 j_\ell(ks) \xi_{\ell'}(s) Q_L(s) \ ,
\end{equation}
where we have defined the window functions
\begin{equation}\label{eq:Q_L}
    Q_L(s) \equiv \frac{(2L+1)}{N_P}\int \frac{d\Omega_s}{4\pi} \int d^3 r_1 \Theta(\r_1) \Theta(\r_1 + \s) \bar n_w(\r_1) \bar n_w(\r_1 + \s)\mathcal{L}_{L}(\mu) \ .
\end{equation}
Inserting the relation between the multipoles of the correlation function and those of the  power spectrum, 
\begin{equation}\label{eq:xi_from_ps}
    \xi_{\ell'}(s) = i^{\ell'} \int \frac{dk'}{2\pi^2} k'^2 \, P_{\ell'}(k') j_{\ell'}(k's) \ , 
\end{equation}
we finally obtain
\begin{equation}\label{eq:P_W}
\braket{\hat P_\ell(k)} = \int dk' \ k'^2 \sum_{\ell'} W_{\ell \ell'}(k, k') P_{\ell'}(k') \ ,
\end{equation}
where we have defined
\begin{equation}\label{eq:W}
    W_{\ell, \ell'}(k, k') = \frac{2}{\pi} (2\ell+1) (-i)^\ell i^{\ell'} \int ds \ s^2 j_\ell(ks) j_{\ell'}(k's) \sum_L \left(\begin{matrix}
\ell & L & \ell'\\
0 & 0 & 0
\end{matrix}\right)^2 Q_L(s) \ .
\end{equation}
\end{widetext}
Notice that, for clarity, we have neglected the integral constraints~\cite{deMattia:2019vdg}, as well as wide-angle contributions~\cite{Beutler:2018vpe}.~\footnote{We have checked that neglecting the integral constraints in the BOSS full-shape analysis leads to small shifts in the posteriors of $\lesssim 1/4 \cdot \sigma$.} 
Our master formula is~Eq.~\eqref{eq:P_W}: to predict the observed power spectrum $\braket{\hat P_\ell(k)}$, we simply need to convolve our predictions $P_{\ell'}(k')$ with $W_{\ell, \ell'}(k,k')$ given by Eq.~\eqref{eq:W}. 
$W_{\ell, \ell'}(k,k')$ can be precomputed, and the only input we need is $Q_L(s)$. 

The window function $Q_L(s)$, Eq.~\eqref{eq:Q_L}, can be obtained in the following way~\cite{Beutler:2018vpe}. 
Using Eq.~\eqref{eq:xi_from_ps} and the identity
\begin{equation}
\int dk \frac{(ks)^2}{2\pi^2} j_L(ks) j_L(ks') = \frac{1}{4\pi}\delta_D(s-s') \ ,
\end{equation}
where $\delta_D$ is the Dirac delta distribution, 
we see that
\begin{equation}\label{eq:Q_L_s_from_Q_L_k}
    Q_L(s) = i^L  \int \frac{dk}{2\pi^2} k^2 \mathcal{Q}_L(k) j_L(ks)\ ,
\end{equation}
where $\mathcal{Q}_L(k)$ is the expectation value of a power spectrum as defined in Eq.~\eqref{eq:mean_fkp_estimator} given $\xi(\s, \r_1) \equiv 1$. 
Therefore, $\mathcal{Q}_L(k)$ can be measured as the power spectrum $P^r_L(k)$ of random objects (whose distribution is approaching Poisson) within the same geometry survey that we are dealing with,
\begin{equation}\label{eq:Q_L_k}
    \mathcal{Q}_L(k) \equiv \alpha \braket{\hat P^r_L(k)} \ , 
\end{equation}
where $\alpha=N_g/N_r$ is the ratio of the number of data ``galaxy'' objects to the number of random objects.
Such catalog of random objects is already available to us, as it is also required for the estimation of the power spectrum. 

The key point is the following: 
$\mathcal{Q}_L(k)$ is normalized by the same normalization factor as $P_\ell(k)$, namely, $N_P$. 
As such, in the limit of vanishing separation $s\rightarrow 0$, the window function monopole does not go to unity, $Q_0(s) \neq 1$, but instead 
\begin{equation}\label{eq:Q_0}
    Q_0(s\rightarrow 0) \rightarrow \frac{1}{N_P}\int d^3 r_1 \bar n_w^2(\r) \ . 
\end{equation}
Given that one does not know the value of the numerator in the equation above prior to making the measurement, $N_P$ can only be estimated \emph{approximately} in order to have $Q_0(s)$ approaching $1$ at vanishing separation $s\rightarrow 0$. 
It is in this sense that $N_P$ is chosen by the user. However, the normalization choice is not important as long as the window function measurements are consistently normalized with the power spectrum measurements. 
Given the measurement protocol sketched above, this is automatic if one is able to evaluate \eqref{eq:Q_L_s_from_Q_L_k} accurately.~\footnote{At~\url{https://github.com/pierrexyz/fkpwin}, we provide a code written to perform the window function measurements, based on \code{nbodykit}.
Let us note that we find that it is not straightforward to get a precise measurements of $\hat{\mathcal{Q}}_L(k)$, namely, the power spectrum of the random objects over the \emph{whole} range of $k$ for which $\hat{\mathcal{Q}}_L(k)$ contributes significantly to the integral in Eq.~\eqref{eq:Q_L_s_from_Q_L_k}. 
Furthermore, the estimator in Eq.~\eqref{eq:Q_L_k} might have a non-negligible variance, given that only one catalog is used. 
We nevertheless have checked that, letting the normalization of the window functions to be different from the one of the power spectrum by a few percents leads to tolerable shifts in the posteriors ($\lesssim 1\sigma/5$) inferred fitting BOSS data. 
For future large-volume datasets, it would be, however, desirable to have a better numerical control over the measurements of $Q_L(s)$ such that the normalization consistency with $P_\ell(k)$ is achieved to sufficient accuracy given increasing precision of the data. } 

In past BOSS full-shape analyses, \emph{e.g.}~\cite{DAmico:2019fhj,Ivanov:2019pdj,Colas:2019ret,Ivanov:2020ril,DAmico:2020ods}, the window function normalizations were instead inconsistently enforced to $Q_0^{\rm wrong}(0) \equiv 1$, while in reality $Q_0(0) \sim 0.9$ given the choice of $N_P$. 
Such inconsistency of $\sim 0.9$ led to a shift in $A_s$ of around $- 1\sigma$ depending on the normalization choice. 
Let us list two choices for the normalization factor $N_P$: 
\begin{itemize}
    \item \textit{Choice 1}: $N_P = \alpha \sum_{\lbrace i \in \text{randoms}\rbrace} \bar n(\r_i) w^2_{\rm FKP}(\r_i)$.~\footnote{Naively one might think that the sum over enough objects is a good approximation to the volume integral; 
    Actually, \textit{Choice 1} poorly estimates the integral in Eq.~\eqref{eq:Q_0} because in the FKP estimator, $\bar n$ is measured from the grid for FFT with finite cell resolution, while in \textit{Choice 1}, we are counting the objects instead.} 
    This was the choice in Ref.~\cite{BOSS:2016psr}, which measurements were used in, \emph{e.g.}, Refs.~\cite{Ivanov:2019pdj,Ivanov:2020ril}. 
    \item \textit{Choice 2}: $N_P = \ \mathcal{A} * \int dr \bar n_w^2(r)$, where $\bar n_w(r)$ is inferred from counting galaxies and binning them in shells and $\mathcal{A}$ is an associated estimated area.~\footnote{We thank Hector Gil-Mar\'in for private correspondence on this point.} 
    This was the choice in Ref.~\cite{Gil-Marin:2015sqa}, which measurements $\mathcal{P}_\textsc{fkp}^\textsc{lz/cm}$ were used in, \emph{e.g.}, Refs.~\cite{DAmico:2019fhj,Colas:2019ret,DAmico:2020ods}. 
    $\mathcal{P}_\textsc{fkp}^\textsc{lz/cm}$, as defined in Table~\ref{tab:twopoint_summary}, is assigned window functions that are inconsistently normalized. 
\end{itemize}
We stress again that those choices are not important as long as the same $N_P$ is used to normalize the window functions and the power spectrum measurements. 
As already mentioned in the main text, except for $\mathcal{P}_\textsc{fkp}^\textsc{lz/cm}$ that is used in this paper for illustration purposes, all power spectrum measurements obtained with the FKP estimator, namely, $P_\textsc{fkp}^\textsc{lz/cm}$ and $P_\textsc{fkp}^{z_1/z_3}$, are instead consistently normalized with their window functions (see Table~\ref{tab:twopoint_summary} for more details on the measurements). 
We finish this section by noting that, in analyses using measurements obtained from the FKP estimator, but also from the other estimators, the posteriors may depend on the effective-redshift approximation used above.
This suggests that, for each estimator, more work is needed to understand the accuracy of this approximation, along the line of, \emph{e.g.},~\cite{Zhang:2021uyp} for the correlation function. \\

\begin{figure*}
\centering
\includegraphics[width=1.82\columnwidth]{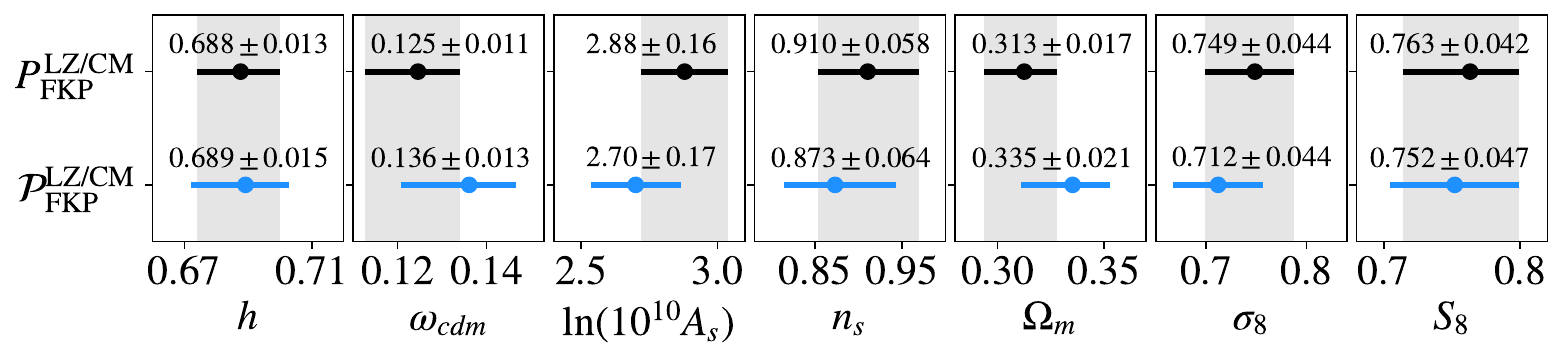}
\caption{
Comparison of $\Lambda$CDM results from BOSS full-shape analysis of the power spectrum measurements $\mathcal{P}_\textsc{fkp}^\textsc{lz/cm}$ and $P_\textsc{fkp}^\textsc{lz/cm}$, analyzed with window functions  inconsistently and consistently normalized, respectively 
(see Tab.~\ref{tab:twopoint_summary}).
The gray bands are centered on the results from the $P_\textsc{fkp}^\textsc{lz/cm}$ data.}
\label{fig:inconsist_norm}
\end{figure*}

In Fig.~\ref{fig:inconsist_norm}, we show a comparison of the 1D posteriors from the full-shape analysis of the BOSS power spectrum measured with the FKP estimator, using window functions with consistent or inconsistent normalization. 
The inconsistency leads to a lower amplitude $A_s$, or equivalently $\sigma_8$, as well as higher $\Omega_m \sim f$, where $f$ is the logarithmic growth rate, through anticorrelation. 
We find notable shifts on $\omega_{\rm cdm}$, $\ln(10^{10}A_s)$, $\Omega_m$ and $\sigma_8$ of 0.9$\sigma$, 1.1$\sigma$, 1.1$\sigma$, and 0.8$\sigma$, respectively.

\section{Additional comparison between the PyBird and CLASS-PT likelihood in EDE}
\label{app:classpt_vs_pybird_EDE}

\begin{figure*}
    \centering
    \includegraphics[width=1.18\columnwidth]{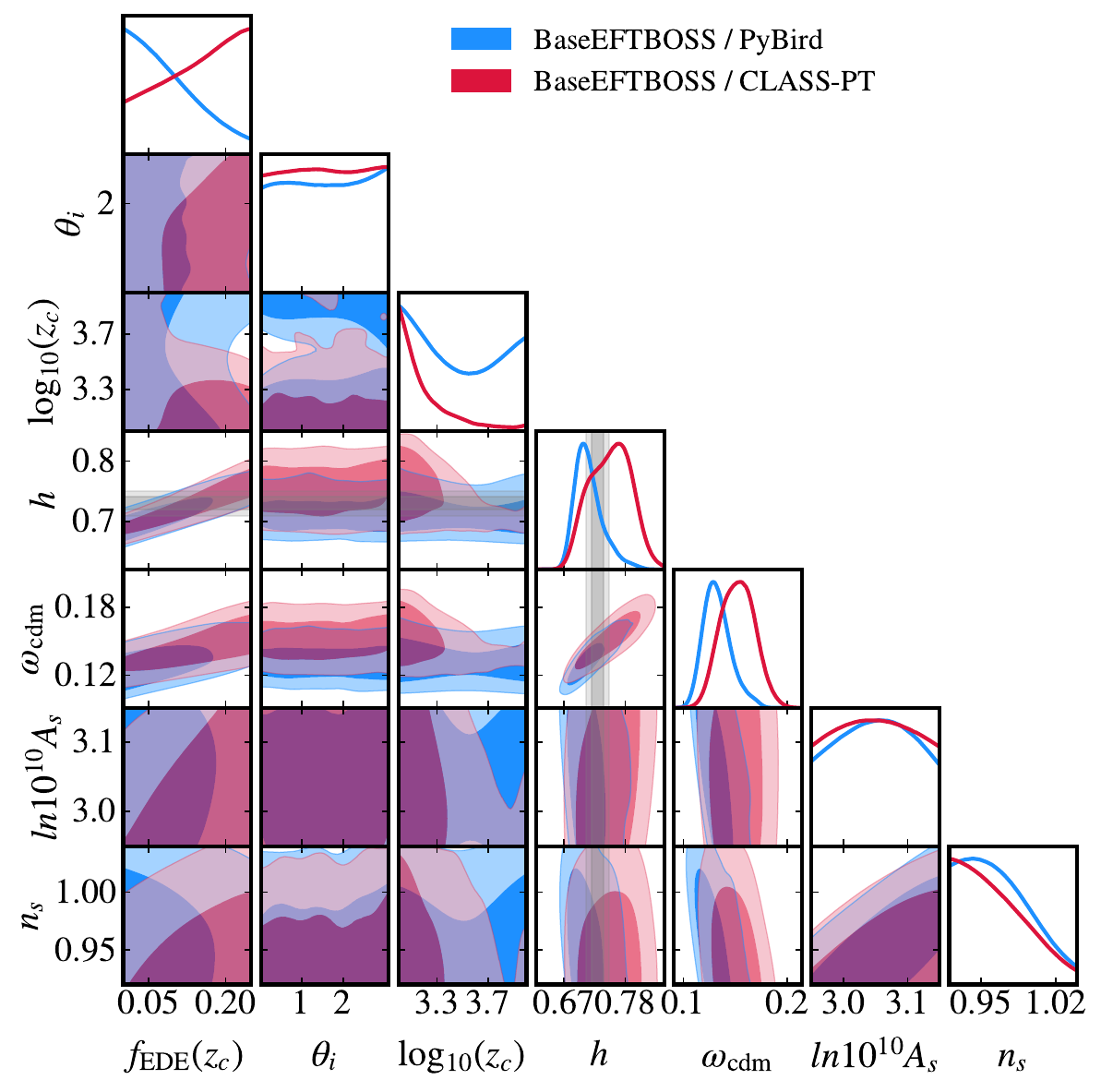}
    \caption{Comparison between the 2D posterior distributions of a subset of parameters in the EDE model reconstructed from the \code{PyBird} or \code{CLASS-PT} likelihood, in combination with BBN+Lens+BAO+Pan18 (\emph{i.e.}, BaseEFTBOSS). }
    \label{fig:EDE_PyBird_vs_CLASSPT}
\end{figure*}

\begin{figure*}
    \centering
    \includegraphics[width=1.18\columnwidth]{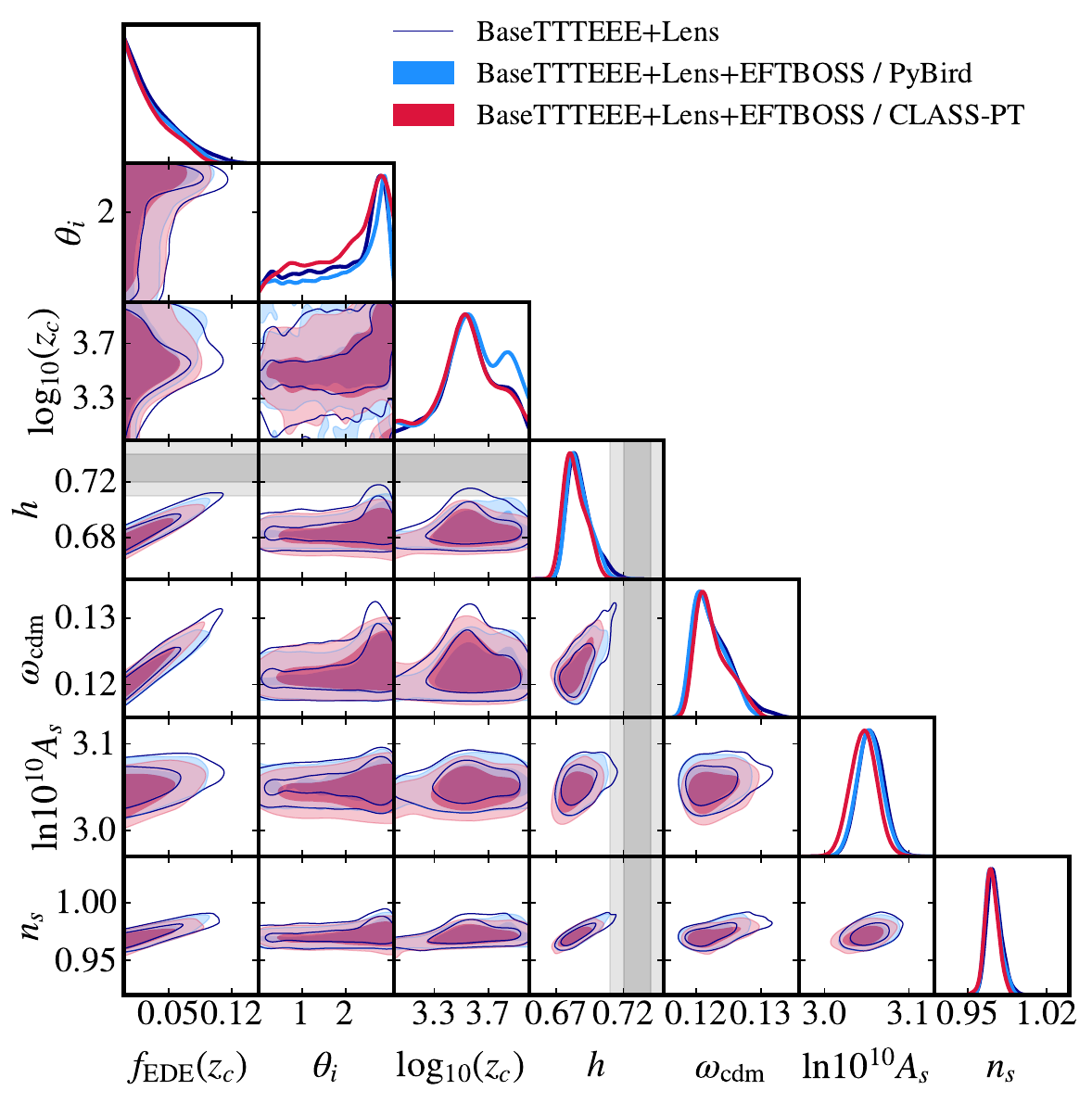}
    \caption{Comparison between the 2D posterior distributions of a subset of parameters in the EDE model reconstructed from the \code{PyBird} or \code{CLASS-PT} likelihood, in combination with BaseTTTEEE+Lens.
    }
    \label{fig:EDE_Planck_PyBird_vs_CLASSPT}
\end{figure*}

\begin{figure*}
    \centering
    \includegraphics[width=1.18\columnwidth]{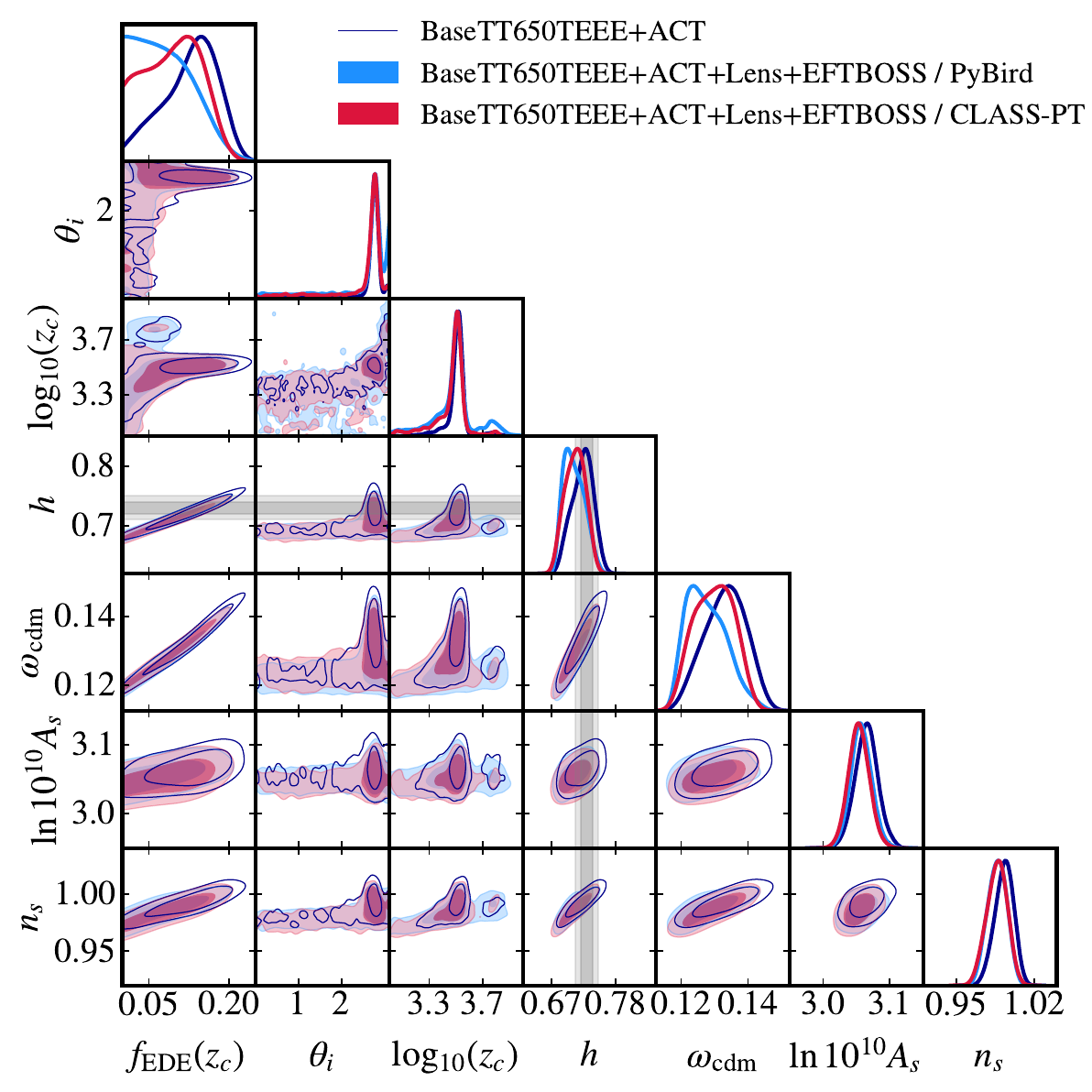}
    \caption{Comparison between the 2D posterior distributions of a subset of parameters in the EDE model reconstructed from the \code{PyBird} or \code{CLASS-PT} likelihood, in combination with BaseTT650TEEE+ACT+Lens.}
    \label{fig:EDE_ACT_noLens_PyBird_vs_CLASSPT}
\end{figure*}

In Figs.~\ref{fig:EDE_PyBird_vs_CLASSPT}, \ref{fig:EDE_Planck_PyBird_vs_CLASSPT}, and \ref{fig:EDE_ACT_noLens_PyBird_vs_CLASSPT}, we show the 2D posterior distributions reconstructed from BaseEFTBOSS, BaseTTTEEE+Lens+EFTBOSS, and  BaseTT650TEEE+ACT+Lens+EFTBOSS, respectively, comparing the results from the \code{PyBird} and the \code{CLASS-PT} likelihoods.~\footnote{For this comparison, LOWZ SGC is not included in the \code{PyBird} likelihood. As expected, we have checked that the addition of this sky cut does not change the posteriors for the corresponding analyses. }
In addition, we recall that EFTBOSS corresponds to $P_\textsc{fkp}^\textsc{lz/cm} + \alpha^{z_1/z_3}_\text{rec}$ in the framework of the \code{PyBird} likelihood and to $P_\textsc{quad}^{z_1/z_3} + \beta^{z_1/z_3}_\text{rec}$ in the framework of the \code{CLASS-PT} likelihood (see Tab.~\ref{tab:twopoint_summary}).
The most striking differences occur in the BaseEFTBOSS alone case, for which \code{CLASS-PT} leads to much weaker constraints on $f_{\rm EDE}(z_c)$ and much larger error bars on $h$ and $\omega_{\rm cdm}$.
The origin of these differences can be traced back to the discussion presented in our companion paper~\cite{Simon:2022lde}, namely, to the choice of the power spectrum estimators, the BOSS post-reconstructed measurements used, the scale cut, the number of multipoles, and more importantly, the choice of EFT parameter priors.
Once {\Planck}TTTEEE or {\Planck}TT650TEEE+ACT data are included in the analysis, we find that the reconstructed posteriors are very similar between the two EFTBOSS implementations and mostly driven by CMB data. We conclude that the main results of this paper, drawn from the combination of CMB and LSS data, are unaffected by the choice of EFT implementation. However, parameter reconstruction based on EFTBOSS data alone may vary at the $1\sigma$ level.

\section{$\chi^2$ per experiment}
\label{app:chi2}

In this appendix, we report the best-fit $\chi^2$ per experiment for both $\Lambda$CDM and EDE models. In Tabs.~\ref{tab:chi2_Planck_LCDM} and \ref{tab:chi2_Planck_EDE} are presented the runs including {\it Planck} data, in Tab.~\ref{tab:chi2_ACT} the runs including ACT data, and in Tab.~\ref{tab:chi2_FullPlanckACT} the combination of the full {\it Planck} data and ACT data. Finally, Tab.~\ref{tab:chi2_PanPlus} present runs including the PanPlus data.

\begin{table*}[t!]
\def\arraystretch{1.2}
\scalebox{1}{
\begin{tabular}{|l|c|c|c|c|c|c|}
    \hline
    \multicolumn{7}{|c|}{$\Lambda$CDM} \\
    \hline
    \hline
    {\emph{Planck}}~high$-\ell$ TTTEEE &2342.2 & 2345.0  &2342.2 & 2344.6&2342.2& 2345.2 \\
    {\emph{Planck}}~low$-\ell$ TT  & 23.4& 22.9 & 23.5&23.0  & 23.4&  22.8\\
    {\emph{Planck}}~low$-\ell$ EE &396.3 & 397.2 & 396.1& 397.2 & 396.3 & 397.2 \\
    {\emph{Planck}}~lensing &8.9 & 9.4 & 9.0 &9.4  &  9.0 & 9.4 \\
    BOSS BAO low$-z$& 1.2& 1.9 &1.2 & 1.8  &  1.2 & 1.9\\ 
    BOSS BAO DR12& 4.3& 3.4 &$-$  & $-$& $-$&$-$ \\ 
    BOSS BAO/$f\sigma_8$ DR12&$-$ &$-$  &6.7 &  5.9& $-$&$-$ \\
    EFTBOSS CMASS &$-$ & $-$ &$-$ &$-$  & 84.6&83.1 \\
    EFTBOSS LOWZ&$-$ & $-$ & $-$&$-$  &  33.5& 33.7  \\
    Pantheon &1027.2 & 1026.9 & 1027.2& 1026.9 & 1027.2 & 1026.9\\
    SH0ES & $-$&  19.9& $-$&  20.4& $-$ &19.8   \\
    \hline
    total $\chi^2_{\rm min}$ & 3803.6& 3826.6 &3805.7 &  3829.1 & 3917.4&3940.0 \\
    \hline
    $Q_{\rm DMAP}$ & \multicolumn{2}{|c|}{4.8$\sigma$}&\multicolumn{2}{|c|}{4.8$\sigma$} &\multicolumn{2}{|c|}{4.8$\sigma$}\\
    \hline
\end{tabular}}
\caption{Best-fit $\chi^2$ per experiment (and total) for $\Lambda{\rm CDM}$ when fit to different data combinations: BaseTTTEEE+Lens, BaseTTTEEE+Lens+$f\sigma_8$, BaseTTTEEE+Lens+EFTBOSS, with and without SH0ES. We also report the tension metric $Q_{\rm DMAP}\equiv \sqrt{\chi^2({\rm w/~SH0ES})-\chi^2({\rm w/o~SH0ES})}$.  }
\label{tab:chi2_Planck_LCDM}
\end{table*}

\begin{table*}[t!]
\def\arraystretch{1.2}
\scalebox{1}{
\begin{tabular}{|l|c|c|c|c|c|c|c|}
    \hline
     \multicolumn{7}{|c|}{EDE} \\
    \hline
    \hline
    {\emph{Planck}}~high$-\ell$ TTTEEE  & 2339.4& 2341.5 & 2339.1& 2340.9 & 2339.3& 2341.1\\
    {\emph{Planck}}~low$-\ell$ TT  & 21.8 & 20.4 & 22.0 & 20.6 &  21.1& 20.5 \\
    {\emph{Planck}}~low$-\ell$ EE  & 396.4  &396.8   & 396.1&  396.4& 396.1 & 396.9  \\
    {\emph{Planck}}~lensing  & 9.5& 10.0 &9.3 & 9.9 & 9.6 &  9.9\\
    BOSS BAO low$-z$  & 1.6&1.8  & 1.4&  1.7&  1.4 & 1.9 \\ 
    BOSS BAO DR12 & 3.7&3.5  & $-$& $-$ & $-$& $-$\\ 
    BOSS BAO/$f\sigma_8$ DR12 &$-$ &$-$  &6.5& 7.0 &$-$ &$-$\\
    EFTBOSS CMASS&$-$ &$-$  & $-$&  $-$&84.1 & 83.3 \\
    EFTBOSS LOWZ & $-$&$-$  &$-$ &$-$  & 34.0 & 34.4 \\
    Pantheon & 1027.0& 1026.9 & 1027.0&1026.9  & 1027.0 & 1026.9\\
    SH0ES & $-$&2.0  &$-$ & 3.2 & $-$ & 2.3 \\
    \hline
    total $\chi^2_{\rm min}$ & 3799.2&3802.9  & 3801.8& 3806.1 & 3912.7 & 3917.3\\
    $\Delta \chi^2_{\rm min}({\rm EDE}-\Lambda{\rm CDM})$ & -3.8 & -23.7 & -3.9&-23.0  & -4.7 & -22.7\\
        Preference over $\Lambda$CDM & 1$\sigma$& 4.2$\sigma$&  1.1$\sigma$& 4.1$\sigma$ & 1.3$\sigma$ & 4.1$\sigma$  \\
    \hline
    $Q_{\rm DMAP}$&\multicolumn{2}{|c|}{1.9$\sigma$} &\multicolumn{2}{|c|}{2.0$\sigma$}& \multicolumn{2}{|c|}{2.1$\sigma$}\\
    \hline
\end{tabular}}
\caption{Best-fit $\chi^2$ per experiment (and total) for EDE when fit to different data combinations: BaseTTTEEE+Lens, BaseTTTEEE+Lens+$f\sigma_8$, BaseTTTEEE+Lens+EFTBOSS, with and without SH0ES. We also report the $\Delta \chi^2_{\rm min}\equiv\chi^2_{\rm min}({\rm EDE})-\chi^2_{\rm min}(\Lambda{\rm CDM})$ and the tension metric $Q_{\rm DMAP}\equiv \sqrt{\chi^2({\rm w/~SH0ES})-\chi^2({\rm w/o~SH0ES})}$.  }
\label{tab:chi2_Planck_EDE}
\end{table*}

\begin{table*}[t!]
\def\arraystretch{1.2}
\scalebox{1}{
\begin{tabular}{|l|c|c|c|c|c|c|c|c|c|c|}
    \hline
    & \multicolumn{5}{|c|}{$\Lambda$CDM}& \multicolumn{5}{|c|}{EDE} \\
    \hline
    \hline
    {\emph{Planck}}~high$-\ell$ TT650TEEE &1843.5 & 1842.6 & 1842.9 & 1842.8&1842.6 & 1837.5 & 1838.0 & 1836.9 & 1836.8 & 1837.7 \\
    {\emph{Planck}}~low$-\ell$ TT & 21.5  & 21.7& 21.5&  21.7& 21.8 & 20.7&  20.9& 20.8& 20.9 & 21.2\\
    {\emph{Planck}}~low$-\ell$ EE &395.7  &395.7 &  395.8 & 395.9 & $-$ & 395.8&  395.8& 395.8& 395.8 & 395.8 \\
    {\emph{Planck}}~lensing  & $-$ & $-$ & $-$  & 9.0 &9.0 & $-$  &  $-$ & $-$& 10.2 &9.9 \\
    ACT DR4& 293.8 &294.5 & 294.4 & 294.2&294.3 & 285.4&285.0& 285.9& 286.4 & 286.9\\
    BOSS BAO low$-z$&  1.5& 1.4 & 1.6 & 1.5&1.4 &2.1  & 2.0&  2.4& 2.3 &1.9 \\    
    BOSS BAO DR12&3.7 &$-$ & $-$& $-$ & $-$ & 3.5&$-$ & $-$ &  $-$ & $-$\\
    BOSS BAO/$f\sigma_8$ DR12& $-$ & 6.1 & $-$& $-$ & $-$&$-$ & 7.2 & $-$ &  $-$ & $-$\\
    EFTBOSS CMASS & $-$ &  $-$  & 83.4 & 83.6 & 84.9 & $-$ &  $-$  &  84.5& 84.3 &84.3\\
    EFTBOSS LOWZ &$-$  & $-$ & 33.7 &33.7& 33.7& $-$  &  $-$  & 35.1&34.7 &34.4 \\
    Pantheon&1026.8 & 1027.0 & 1027.0 & 1027.0& $-$& 1026.9 & 1026.9  & 1026.9& 1026.9 & $-$\\
        Pantheon+ & $-$ & $-$& $-$&$-$ & 1411.8&$-$ &$-$ & $-$ & $-$ & 1413.0  \\

    \hline
    total $\chi^2_{\rm min}$ &3586.5 & 3589.1& 3700.3 & 3709.5 &4094.3 & 3571.9 & 3575.8 & 3688.3 & 3698.4 & 4085.1\\
    $\Delta \chi^2_{\rm min}({\rm EDE}-\Lambda{\rm CDM})$ & $-$ & $-$& $-$ & $-$ & $-$ & -14.6 & -13.3 & -12.0 & -11.1 & -9.2 \\
    Preference over $\Lambda$CDM & $-$ & $-$& $-$ & $-$& $-$ & 3.1$\sigma$ & 2.9$\sigma$& 2.7$\sigma$& 2.5$\sigma$ & $2.2\sigma$ \\
    \hline
\end{tabular}}
\caption{Best-fit $\chi^2$ per experiment (and total) for $\Lambda{\rm CDM}$ and EDE when fit to different data combinations: BaseTT650TEEE+ACT, BaseTT650TEEE+ACT+$f\sigma_8$, BaseTT650TEEE+ACT+EFTBOSS, BaseTT650TEEE+ACT+Lens+EFTBOSS, and BaseTT650TEEE+ACT+Lens+EFTBOSS+PanPlus. We also report the $\Delta \chi^2_{\rm min}\equiv\chi^2_{\rm min}({\rm EDE})-\chi^2_{\rm min}(\Lambda{\rm CDM})$ and the corresponding preference over $\Lambda$CDM, computed assuming the $\Delta\chi^2$ follows a $\chi^2$ distribution with three degrees of freedom.}
\label{tab:chi2_ACT}
\end{table*}

\begin{table*}[t!]
\def\arraystretch{1.2}
\scalebox{1}{
\begin{tabular}{|l|c|c|c|c|}
    \hline
    & \multicolumn{2}{|c|}{$\Lambda$CDM}& \multicolumn{2}{|c|}{EDE} \\
    \hline
    \hline
    {\emph{Planck}}~high$-\ell$ TTTEEE  & 2349.8 & 2352.0& 2346.2& 2347.2 \\
    {\emph{Planck}}~low$-\ell$ TT & 22.4 & 22.0& 21.9&21.2 \\
    {\emph{Planck}}~low$-\ell$ EE &396.2 & 396.8&396.1 & 396.4 \\
    {\emph{Planck}}~lensing  &8.9 & 8.9& 9.6&9.8 \\
    ACT DR4 & 240.6& 241.0& 236.8& 236.2 \\
    BOSS BAO low$-z$& 1.4&2.0 &1.7 & 2.2\\   
    EFTBOSS CMASS &84.1 &82.9 &84.2 & 84.2 \\
    EFTBOSS LOWZ  & 33.6 &33.8 & 34.2& 34.6 \\
    Pantheon &1027.1 &1026.9 &1026.9 & 1026.9 \\
    SH0ES &$-$ & 19.5& $-$ & 1.10\\
    \hline
    total $\chi^2_{\rm min}$ &4164.0 &4185.9 & 4157.6 & 4159.8  \\
    $\Delta \chi^2_{\rm min}({\rm EDE}-\Lambda{\rm CDM})$ & $-$ & $-$ &-6.4 &-26.1 \\
        Preference over $\Lambda$CDM &$-$& $-$ &  1.7$\sigma$ & 4.4$\sigma$\\
       \hline
    $Q_{\rm DMAP}$&\multicolumn{2}{|c|}{4.7$\sigma$} &\multicolumn{2}{|c|}{1.5$\sigma$}\\
   \hline
\end{tabular}}
\caption{Best-fit $\chi^2$ per experiment (and total) for $\Lambda{\rm CDM}$ and EDE when fit to BaseTTTEEE+ACT+Lens+EFTBOSS, with and without SH0ES. We also report the $\Delta \chi^2_{\rm min}\equiv\chi^2_{\rm min}({\rm EDE})-\chi^2_{\rm min}(\Lambda{\rm CDM})$ and the tension metric $Q_{\rm DMAP}\equiv \sqrt{\chi^2({\rm w/~SH0ES})-\chi^2({\rm w/o~SH0ES})}$.  }
\label{tab:chi2_FullPlanckACT}
\end{table*}

\begin{table*}[t!]
\def\arraystretch{1.2}
\scalebox{1}{
\begin{tabular}{|l|c|c|c|c|}
    \hline
    & \multicolumn{2}{|c|}{$\Lambda$CDM}& \multicolumn{2}{|c|}{EDE} \\
    \hline
    \hline
    {\emph{Planck}}~high$-\ell$ TTTEEE  & 2346.18 & 2349.5& 2344.0& 2346.9 \\
    {\emph{Planck}}~low$-\ell$ TT &23.0 & 22.4& 22.3& 21.0 \\
    {\emph{Planck}}~low$-\ell$ EE & 396.1 &  397.7&396.3 &  396.3\\
    {\emph{Planck}}~lensing  & 8.8& 9.1&9.0 & 9.6\\
    BOSS BAO low$-z$&1.1 & 2.1&1.3 & 1.8\\   
    EFTBOSS CMASS &85.2 & 82.9&85.0 & 85.1  \\
    EFTBOSS LOWZ  &33.6 & 33.8& 33.8&  34.6\\
    Pantheon+ &1411.1 & $-$&1411.6 &  $-$\\
        Pantheon+SH0ES & $-$ &1321.9 & $-$&  1291.6\\

    \hline
    total $\chi^2_{\rm min}$ & 4305.1&4219.3 &4303.2 & 4187.0  \\
    $\Delta \chi^2_{\rm min}({\rm EDE}-\Lambda{\rm CDM})$ & $-$ &  $-$ &-1.9 & -32.3\\
    Preference over $\Lambda$CDM &$-$& $-$ &  0.5$\sigma$ & 5$\sigma$\\
   \hline
\end{tabular}}
\caption{Best-fit $\chi^2$ per experiment (and total) for $\Lambda{\rm CDM}$ and EDE when fit to BaseTTTEEE+Lens+EFTBOSS+PanPlus, with and without SH0ES. We also report the $\Delta \chi^2_{\rm min}\equiv\chi^2_{\rm min}({\rm EDE})-\chi^2_{\rm min}(\Lambda{\rm CDM})$ and the corresponding preference over $\Lambda$CDM, computed assuming the $\Delta\chi^2$ follows a $\chi^2$ distribution with three degrees of freedom.}
\label{tab:chi2_PanPlus}
\end{table*}

\newpage
\bibliography{biblio}

\end{document}